\documentclass[journal=jctcce,manuscript=article,layout=traditional]{achemso} 

\PassOptionsToPackage{letterpaper,top=2cm,bottom=2cm,left=3cm,right=3cm,marginparwidth=1.75cm}{geometry}

\usepackage[english]{babel}

\usepackage{geometry}

\usepackage{amsmath,bm}
\usepackage{graphicx}
\usepackage[colorlinks=true, allcolors=blue]{hyperref}
\usepackage{amssymb}
\usepackage{mathrsfs}
\usepackage{xcolor}
\usepackage{xfrac}
\usepackage{physics}
\usepackage{comment}
\usepackage{float}
\usepackage{subfig}
\usepackage{bm}
\usepackage[cal=dutchcal]{mathalfa} 

\usepackage{natbib}

\newcommand{\hr}{{\hat{r}}}
\newcommand{\hR}{{\hat{R}}}
\newcommand{\hP}{{\hat{P}}}

\newcommand{\hp}{{\hat{p}}}
\newcommand{\hH}{{\hat{H}}}

\newcommand{\bmu}{{\bm{\mu}}}
\newcommand{\bmm}{{ \bm{\mathcal{m}}}}
\newcommand{\bGamma}{{\bm{\Gamma}}}
\newcommand{\bd}{{\bm{d}}}
\newcommand{\bR}{{\bm{R}}}
\newcommand{\bP}{{\bm{P}}}
\newcommand{\bK}{{\bm{K}}}
\newcommand{\bW}{{\bm{W}}}
\newcommand{\bY}{{\bm{Y}}}
\newcommand{\br}{{\bm{r}}}

\newcommand{\bhR}{\hat{\bm{R}}}
\newcommand{\bhP}{\hat{\bm{P}}}

\newcommand{\mhb}{{\hat {\bm{\mu}}}}
\newcommand{\mmhb}{{\hat {\bm{\mathcal{m}}}}}

\newcommand{\pp}[2]{\frac{\partial {#1}}{\partial {#2}}}

\author{Titouan Duston}
\affiliation{Department of Chemistry, University of Pennsylvania, Philadelphia, Pennsylvania 19104, USA.}
\author{Zhen Tao}
\affiliation{Department of Chemistry, University of Pennsylvania, Philadelphia, Pennsylvania 19104, USA.}
\author{Xuezhi Bian}
\affiliation{Department of Chemistry, University of Pennsylvania, Philadelphia, Pennsylvania 19104, USA.}
\author{Mansi Bhati}
\affiliation{Department of Chemistry, University of Pennsylvania, Philadelphia, Pennsylvania 19104, USA.}
\author{Jonathan Rawlinson}
\affiliation{Department of Mathematics, Nottingham Trent University, Nottingham, UK}
\author{Robert G. Littlejohn}
\affiliation{Department of Physics, University of California, Berkeley, California 94720, USA}
\author{Zheng Pei}
\affiliation{Department of Chemistry, Norman, The University of Oklahoma, Oklahoma, 73104, USA}
\author{Yihan Shao}
\affiliation{Department of Chemistry, Norman, The University of Oklahoma, Oklahoma, 73104, USA}
\author{Joseph E. Subotnik}
\email{subotnik@sas.upenn.edu}
\affiliation{Department of Chemistry, University of Pennsylvania, Philadelphia, Pennsylvania 19104, USA.}

\title[]{A Phase Space Approach to Vibrational Circular Dichroism}

\begin{document}
\maketitle

\begin{abstract}
We show empirically that a phase-space non-Born-Oppenheimer electronic Hamiltonian approach to quantum chemistry (where the electronic Hamiltonian is parameterized by both nuclear position and momentum, $\hH_{PS}(\bR,\bP)$)  is both a practical and accurate means to recover vibrational circular dichroism spectra. We further hypothesize that such a phase space approach may lead to very new dynamical physics beyond spectroscopy circular dichroism, with potential implications for understanding chiral induced spin selectivity (CISS), noting that  classical phase space approaches conserve the total nuclear plus electronic momentum, whereas classical Born-Oppenheimer approaches do not (they conserve only the nuclear momentum).

\end{abstract}

\section{Introduction: Chirality and Circular Dichroism}

\noindent
Molecular chirality occurs whenever a system lacks any inversion or mirror symmetry. Such systems have long fascinated chemists, going all the way back to the work of Pasteur more than one hundred fifty years ago.\cite{Pasteur} Chirality continues to be a target of cutting-edge research, with the 2021 Nobel prize in chemistry being awarded for chiral synthesis\cite{NobelList,NobelMacMillan} and the current explosion of interest in chiral induced spin selectivity,\cite{Naaman2012,Naaman2019,Evers2022} the effect whereby electronic conduction through a chiral medium is found to be spin-polarized.
One hypothesis for explaining the chiral induced spin selectivity (CISS) effect is that coupled nuclear-electronic motion transfers angular momentum from ``chiral phonons''\cite{Zhang2015,Bistoni2021} to electronic spin degrees of freedom --- though this hypothesis is unconfirmed and many details remain uncertain.\cite{Das2022,Fransson2023,Kim2023} Nevertheless, what is clear is that the chirality of a molecular system can substantially affect electronic properties in ways we still do not fully understand and cannot control.
%Microscopic symmetries play a crucial role in elucidating dynamic processes, such as the chirality induced spin selectivity (CISS) effects or the dynamics of spin-dependent chiral phonons. 

To date, the standard approach in the literature for identifying and characterizing chiral molecules and materials is to use circular dichroism (CD) spectroscopy, which measures the differential absorption of molecular systems to circularly polarized light. While non-chiral molecules respond equivalently to both left and right circularly polarized light, the response of chiral molecules is reversed based on the enantiomer.\cite{Berova2000} In particular, because they lack inversion or mirror symmetry, chiral molecules display different magneto-electronic response which leads to a small difference in absorption when exposed to left-handed and right-handed circularly polarized light.
Over the last few decades, in order to treat chirality in different regimes, experimentalists have designed various flavors of CD spectroscopy, electronic circular dichroism (ECD),\cite{Warnke2012} vibrational CD (VCD),\cite{Nafie1976} magnetic CD (MCD),\cite{Stephens1970} magnetic vibrational CD (MVCD), X-ray magnetic CD (XMCD), Raman optical activity (ROA), etc. Effectively, one can measure the differential absorption of molecules in many wavelength regimes (from radio-wave to x-ray) and in many environments (e.g., including or not including magnetic fields, solvated and unsolvated). 

In order to interpret the different CD spectroscopies listed above, the job of the theoretical chemist is usually to calculate the rotatory strength $\mathcal{R}$, which is related to the difference in left-handed vs. right-handed absorption when averaged over all incoming directions of light and molecular geometries. Mathematically, this difference in molecular absorption can be quantified using the product of an electronic transition matrix element ($\bm{\mu}$) dotted into a magnetic transition matrix element ($\bm{m}$)
\begin{eqnarray}
\label{eq:Rsimple}
    \mathcal{R} = \text{Im}(\bm{\mu}_{if}\cdot{\bm{\mathcal{m}}}_{fi})
\end{eqnarray}
where $i$ and $f$ represent the initial and final states, before and after absorption of a photon. One can find many review articles discussing the strategies for calculating $\mathcal{R}$ \cite{ECDReview,VCDROAReviewNafie,MCDReview,Stephens1985,nafie1979optical,Nafie1987book} as well as below (Sec. \ref{sec:rotstr}). Historically, one of the difficulties encountered when calculating the rotatory strength is that, because we are limited to a finite basis, magnetic response properties will naively depend on the choice of origin. 
%One of the great achievements as far as calculating magnetic shielding tensors with NMR was the introduction of gauge-invariant atomic orbitals(GIAOs)\cite{pulay} which dramatically improved performance, and GIAOs have also been used quite successfully in the past  to calculate the $\mathcal{R}$.\cite{VCDGIAO}

\subsection{Vibrational Circular Dichroism and Magnetic Field Pertubation Theory}
Now, it is crucial to emphasize that calculating $\mathcal{R}$ for each of the different CD spectroscopies above can require quite different methodologies. In particular, relative to the ECD signal, the calculation of a VCD signal can be much less straightforward because, 
%That being said, it is crucial to recognize that CD spectroscopy comes in two flavors: electronic circular dichroism (ECD) and vibrational circular dichroism (VCD). At their most basic level,  ECD and VCD are similar inasfar as both methods
%Both ECD and VCD signals are theoretically quantified using their rotational strengths, a quantum-mechanical derived quantity. For the most part, computing ECD is straightforward (relatively speaking). The basic idea is that one must look at the response of an electron's degrees of freedom to both electronic and magnetic degrees of freedom. 
%Perhaps the most conceptually difficult component of ECD is that the magnetic transition moment depends on the origin. That being said, a great deal has already been written about how to generate ECD spectra using a series of gauge choices to avoid this problem.\cite{ECDGauge} \\
 within Born-Oppenheimer theory, the electronic contribution to the magnetic transition dipole moment vanishes between any vibrational modes $i$ and $f$; formally $\mathcal{R} = 0$. \cite{cohan1966BO} 
 To see why this is so, note that the BO Hamiltonian is time-reversible so that all nondegenerate eigenstates are also time-reversible (with zero average electronic momentum, $\left<\hat{\bm{p}}\right>=\left<\hat{\bm{m}}\right>=0$). This inability of standard BO theory to calculate VCD spectra stimulated the original work of Nafie to go beyond BO theory and construct the relevant nonzero matrix elements between perturbatively corrected BO states.\cite{Nafie1977BO}  
At present, there are two main approaches: (i) The first and less common approach is invokes a sum over excited states (SOS) \cite{nafie1983SOS}
in order to generate a perturbatively correct mixed electronic-nuclear wavefunction within a basis of BO wavefunctions.
(ii) The second and now more common approach was formulated by Stephens and Buckingham, which computes the orbital response to a magnetic field perturbation (MFP),\cite{Stephens1985,BUCKINGHAM1987} which formally recovers the same sum over excited states but in a more tractable manner.

For VCD (as for most vibrational problems), one quantifies the electric dipole component of Eq. \ref{eq:Rsimple} by expanding the ground-state electronic dipole in nuclear position and then quantizing and taking matrix elements of the nuclear position operator:
\begin{eqnarray}
   \mhb = \bm{\mu}(\bR^{eq})+ \sum_{A \alpha} \frac{\partial \mhb}{\partial R_{A\alpha}}(\hR_{A \alpha} - R^{eq}_{A \alpha}) + ... 
\end{eqnarray}
In principle, one would like to expand and evaluate the magnetic transition moment, $\bm{\mathcal{m}}$, in a similar fashion -- but as just discussed, within the BO approximation, $\bm{\mathcal{m}}=\bm{0}$ (and so are all spatial  derivatives of $\bm{\mathcal{m}}$, $\pp{\bmm}{\bR} = 0$). That being said, the insight of Nafie and Stephens was that, as the nuclei move, that motion induces electronic motion, which should allow an expansion of the magnetic transition moment, $\bm{\mathcal{m}}$, in terms of a nuclear canonical momentum (where ${\bf{P}}^{eq}_{A} = 0$):
\begin{eqnarray}
   \mmhb = \sum_{A \alpha} \frac{\partial \mmhb}{\partial P_{A\alpha}}(\hP_{A \alpha} - P^{eq}_{A \alpha}) + ... 
\end{eqnarray}

 The success of the MFP approach is that one can map $\frac{\partial \bm{\mathcal{m}}}{\partial P}$ to a double response function calculation; see Eq. \ref{eq:dmdpMFPequal} below. Now, in principle, one might ask: why intuitively should it be more difficult or involved to calculate a VCD spectrum relative to IR or ECD spectrum? Why does the calculation of $\frac{\partial \bm{\mathcal{m}}}{\partial P}$ require a double perturbation in Eq. \ref{eq:dmdpMFPequal}, whereas the calculation of $\frac{\partial \bm{\mu}}{\partial R}$ requires only a single perturbation (see Eq. \ref{eq:dmudRhf})? Just as IR spectra can be interpreted by performing a Fourier transform on the dipole-dipole correlation function, can we interpret VCD spectra as evolving from dynamics on a non-BO surface? In short, is there a non-BO approach that can put VCD theory in a less esoteric framework? 

In a sense, Nafie asked these questions long ago, which led him to introduce nuclear velocity perturbation theory, whereby one introduces gauges to the atomic orbitals\cite{Nafie1992} in an attempt to capture how electronic wavefunctions depend self-consistently on nuclear velocity and fashion a more natural treatment of VCD. Unfortunately, the need for multiple orbital response calculations leads to an expensive algorithm, and so the NVP approach has not been studied in as much detail over the years. Nevertheless, Nafie laid the groundwork for future nonadiabatic methods designed to recover VCD spectra. And recently, the community of exact factorization practitioners has taken up the cause and shown progress with NVP.\cite{ditler2022,Sebastiani2013} In this paper, we will take a different approach and show that VCD spectra can easily be interpreted through a simple and intuitive phase-space approach to electronic structure theory.

\subsection{Phase Space Approaches And An Outline of this Paper}\label{sec:Outline}

The concept of a phase-space approach to nuclear dynamics is quite old; for a review of progress, see the work of Micha {\em et al.} \cite{MichaPSSH} The basic idea is to parameterize the electronic Hamiltonian by both position and momentum, $H_{PS}(\bR,\bP)$. In doing so, one can show that, even with purely classical nuclear dynamics, the resulting dynamics recovers total linear and angular momentum conservation;\cite{coraline24gamma} we emphasize that, by contrast, for systems with odd numbers of electrons, classical BO dynamics violates momentum conservation (unless a Berry force is included\cite{berryforce}). 
Moreover, as argued originally by Shenvi and discussed more below, nonadiabatic surface hopping algorithms are most natural with phase-space surfaces.  Indeed, in the context of intersystem crossing (ISC) and/or problems with spin degeneracy,\cite{ISCReview} standard surface hopping must be adjusted because momentum rescaling is not well-defined if the rescaling direction is complex-valued. For  these reasons, phase-space methods represent a very attractive tool to explore electron-phonon couplings, especially in the context of problems with spin dependent Hamiltonians. And yet, to date, there has been little experimental benchmarking of phase-space approaches. (Interestingly, among those chemists and physicists seeking to build a non-Born-Oppenheimer  framework to nonadiabatic dynamics, most of the focus nowadays is on  exact factorization methods\cite{abedi:2010:prl_exact_factorization, abedi:2012:jcp_exact_factorization, agostini:2023:jpcl:ci, maitra:2024:electronic_coherences_exact_factorization, curchod:2016:exact_factorization} -- which are also attractive methods that conserve momentum\cite{li_gross:2022:PRL:angular_momentum_transfer} but algorithmically require tools even farther away from standard BO theory than simple phase space methods.)

With this background in mind, the goal of the present paper is to directly validate a phase-space electronic structure approach by comparing against experimental VCD spectra for a series of small rigid molecules. This article is structured as follows. 
In Sec. \ref{sec:BOapproach} , we review the approximations taken in the standard BO approach to nonadiabatic theory. In Sec. \ref{sec:phasespaceframework}, we review the basic theory of phase-space electronic Hamiltonians, highlighting the essential equations needed for momentum conservation. In Sec. \ref{sec:gamma2}, we describe one second derivative coupling we have recently implemented. In Sec. \ref{sec:PSModes}, we discuss how to construct normal modes in a phase-space framework. In Sec. \ref{sec:rotstr} we discuss the form of the rotational strength in the phase-space framework. Finally, in Sec. \ref{sec:HFexpressions}, we define the transition electric and magnetic dipole moments at our level of theory. Thereafter, in Sec. \ref{sec:Results}, we present our results which strongly validate phase-space theory (at least numerically). Lastly, in Sec. \ref{sec:Conclusion}, we interpret our findings and speculate on their implications for dynamics beyond VCD theory.

Before concluding this introduction, a word about notation is in order. Henceforward,
\begin{itemize}
    \item Roman letters I,J,K denote adiabatic states
    \item  $\left\{\mu, \nu, \lambda, \sigma \right\}$ index atomic orbitals $\left\{\chi_{\mu},\chi_{\nu} ,\chi_\lambda, \chi_\sigma\right\}$
    \item $\left\{\alpha, \beta, \gamma, \delta\right\}$ index the $x,y,z$ Cartesian directions
    \item Greek letters  $\left\{\tau, \kappa, \eta, \theta \right\}$ index spin of molecular orbital coefficients
    \item  ${A, B,...}$ index atomic centers
    \item Bold indicates three-dimensional quantities (operators or c-numbers)
    \item Hats indicate operators (nuclear and electronic).
    \item $\bR_A$ and $\bP_A$ denote the three-dimensional position and momentum for nucleus A, while $\bR$ and $\bP$ denote a 3N-dimensional vector for all nuclear positions and momenta. 
    \item Adiabatic wavefunctions are denoted $\Phi_J$ with corresponding nuclear wavefunction $\Omega_J$.
    \item $k$ indexes normal modes
    \item The charge of an electron is denoted $-e$ (i.e. we fix $e >0$).
\end{itemize}

\section{Theory: A Phase Space Hamiltonian Framework}

\subsection{Standard Born-Oppeneimer Approach to Nonadiabatic Theory}\label{sec:BOapproach}

To understand how phase-space electronic Hamiltonians can be rationalized, consider the standard quantum mechanical molecular Hamiltonian
\begin{eqnarray}
\label{eq:standard}
\hat{H}_{tot} = \hat{T}_{nuc} + \hat{H}_{el}
\end{eqnarray} 
where $ \hat{T}_{nuc}$ is the kinetic nuclear energy operator (that depends on canonical momentum $\bP$) and $\hat{H}_{el}$ is the electronic Hamiltonian (that depends on $\bR$). According to standard Born-Oppenheimer theory, in order to model dynamics, one first diagonalizes the electronic Hamiltonian $\hat{H}_{el}$, 
\begin{eqnarray}
\label{eq:Hel_diag}
\hat{H}_{el} \ket{\Phi_K} = E_K(\bR) \ket{\Phi_K}.
\end{eqnarray} Second, one expresses the combined nuclear-electronic wavefunction $\Psi(\br, \bR)$ in the basis of adiabatic states:
\begin{eqnarray}
\Psi(\br, \bR) = \sum_J \Phi_J(\bm{r};\bR)\Omega_J(\bR).
\end{eqnarray}

%for simplicity let us consider a single nuclear degree of freedom, wherein $\hat{T}_{nuc}$ can be written as 

%Due to parametric dependency of adiabatic states on R, evaluating $\bra{\Phi_I}\hat{T}_{nuc} \ket{\Phi_J}$ requires the use of the chain rule in the action of the Laplacian. The Hamiltonian can then be written as
%\begin{equation}
%\begin{aligned} 
%\bra{\Phi_i}\hat{T}_{nuc} \ket{\Phi_J} &= \frac{-\hbar^2}{2M} \left[\delta_{IJ} \nabla_R^2 + 2\bra{\Phi_I}\nabla_R\ket{\Phi_J}\nabla_R + \bra{\Phi_i}\nabla_R^2\ket{\Phi_J} \right]\\
% &= \frac{1}{2M} \left[\delta_{IJ} P^2 - 2 i\hbar \bra{\Phi_i}\nabla_R\ket{\Phi_J}P - \hbar^2 \bra{\Phi_i}\nabla_R^2\ket{\Phi_J} \right]
%\end{aligned}
%\end{equation}\
In such a representation, using the fact that $\hat{T}_{nuc} = -\sum_A \frac{\hbar^2}{2M_A} \nabla_{R_A}^2$, it is straightforward to show that the Hamiltonian takes the following form:
\begin{eqnarray}
\label{eq:exactBOH}
    \hat{H}_{tot} = \sum_{IJK} \frac{1}{2M} \ket{\Phi_I} \left( \bm{\hP} \delta_{IJ} - i\hbar \bm{{d}}_{IJ} \right)\cdot
    \left( \bm{\hP} \delta_{JK} - i \hbar \bm{{d}}_{JK} \right)\bra{\Phi_K} + \sum_k E_{KK}(\bhR)\ket{\Phi_K}\bra{\Phi_K}
\end{eqnarray}
where $\bm{\hP} = \frac{\hbar}{i}\frac{\partial}{\partial \bm{R}}$ in the position basis. 
Note that, after an adiabatic transformation of this kind, one has entangled the electronic Hamiltonian and the nuclear kinetic energy. Nevertheless, it is still common to write:
\begin{eqnarray}
    \hat{H}_{tot} = \hat{T}_{nuc}^{ad} +
    \hat{H}_{el}^{ad}
    \label{eq:quickad}
\end{eqnarray}
where, if $\hat{U}$ is the matrix of adiabatic eigenvectors expressed in a diabatic basis, $\hat{T}_{nuc}^{ad} = U^{\dagger} \hat{T}_{nuc} U$ and $\hat{H}_{el}^{ad} = U^{\dagger} \hat{H}_{el} U$.

%For a wavefunction using a non-phase-space BO Hamiltonian $\bra{\Phi_i}\nabla_R\ket{\Phi_i}= 0$, thus $\bra{\Phi_i}\hat{T}_{nuc} \ket{\Phi_i}$ contains only the on-diagonal Born-Oppenheimer correction $\frac{1}{2M} \bra{\Phi_i}\nabla_R^2\ket{\Phi_i}$. Instead, we now choose to include the derivative coupling directly into our electronic Hamiltonian, through a momentum coupling term $2i\hbar \bra{\Phi_i}\nabla_R\ket{\Phi_i} \cdot P $. 
Eq. \ref{eq:exactBOH} is exact (without any approximation). Inevitably, however, in order to simulate a real system, one is forced to make approximations and the most common initial step is to assume that all nuclear motion will be classical. In that vein,
the surface hopping\cite{Tully1990, kapral1999} view of nonadiabatic dynamics is that one interprets dynamics following the Hamiltonian in Eq. \ref{eq:exactBOH} as arising from two steps:
\begin{itemize}
    \item Propagate classical motion along potential energy surface $K$ (as represented by the operator $\hat{H}_{el}^{ad}(\bhR)$).
    \item Hop probabilistically between potential energy surfaces, J and K (as represented by the $\hat{\bP} \cdot \bm{d}_{JK}$ operator).
\end{itemize}
This view of nonadiabatic motion, based on the separation between the $\hat{T}_{nuc}^{ad}$ and $\hat{H}_{el}^{ad}$ terms in Eq. \ref{eq:quickad},
clearly depends delicately on the choice of electronic basis. For the most part, surface hopping is only well-defined in an adiabatic basis -- even though exact quantum dynamics has no preferred basis.\cite{adiabatic}
Nevertheless, this surface hopping approach has been validated for many Hamiltonians (at least for quite a few Hamiltonians).

\subsection{Intuition Behind a Phase-Space Approach}\label{sec:phasespaceframework}

Beginning with Shenvi's seminal work,\cite{shenviPSSH} the original rationale for a phase-space surface hopping formalism was that if one were to partition the Hamiltonian slightly differently from Eq. \ref{eq:quickad},
one could go beyond BO theory by entangling nuclear and electronic motion both through hops {\em and} along dynamics on one surface. To that end, if one is prepared to run a classical simulation, Shenvi proposed rediagonalizing the full Hamiltonian as parameterized by both $\bR$ and $\bP$:
\begin{align}
   \nonumber
    \hat{H}_{Shenvi}(\bR,\bP) & = \sum_{IJKA} \frac{1}{2M_A} \left( \bm{P}_A \delta_{IJ} - i\hbar \bm{{d}}^A_{IJ} \right)\cdot
    \left( \bP_A \delta_{JK} - i \hbar \bm{{d}}^A_{JK} \right)\ket{\Phi_I}\bra{\Phi_K} \nonumber \\
     &  \; \; \; \; \; \; \; \; \; \; \; \; + \sum_k E_{KK}(\bR)\ket{\Phi_K}\bra{\Phi_K} 
    \label{eq:shenvi} \\ 
    \hat{H}_{Shenvi}(\bR,\bP) & \ket{\Psi_{Shenvi}(\bR,\bP)} = E_{Shenvi}(\bR,\bP) \ket{\Psi_{Shenvi}(\bR,\bP)} \label{eq:shenvi_eig}
\end{align}
Note that in Eq. \ref{eq:shenvi}, the nuclear coordinates $\bR$ and $\bP$ are parameters (not operators). As shown in several papers, \cite{shenviPSSH}\cite{izmaylov2016jpc_dboc_pssh} for some model problems, dynamics along the phase-space adiabats in Eq. \ref{eq:shenvi_eig} can produce strong results. Unfortunately, there are several problems:
\begin{enumerate}
    \item The algorithm is not stable. For instance, near a conical intersection, where the derivative coupling diverges, the phase-space adiabatic energies (i.e. the eigenvalues of $\hat{H}_{Shenvi}$) will also diverge.
    \item For problems with spin, the eigenstates in Eq. \ref{eq:Hel_diag} have some an arbitrary gauge (i.e phase). The derivative couplings in Eq. \ref{eq:shenvi} are gauge dependent, and thus the algorithm is not well-defined.
    \item The algorithm in practice is quite expensive because, in order to propagate $\frac{\partial E_{Shenvi}}{\partial \bR}$, one must differentiate the deriative coupling -- which is at least (if not much more) costly than a second-derivative calculation.
\end{enumerate}

For all of these reasons, Shenvi's direct approach has never been used for large or ab initio systems. Nevertheless, we have argued\cite{coraline24gamma} that there is another, simpler phase-space approach which should capture some of the key physics in Eq. \ref{eq:shenvi} but without any of the problems listed above.
The basic idea of our phase-space approach is rooted in the fact that every calculation of the derivative coupling\cite{athavale2023, yarkony} in an atomic orbital basis can always be decomposed into two terms:
\begin{eqnarray}
\label{eq:decompose_d}
    \left<I\middle|\pp{}{R_{A\alpha}}\middle|J \right> = d_{IJ}^{A \alpha} = d_{IJ}^{A \alpha,ETF} + \tilde{d}_{IJ}^{A \alpha}
\end{eqnarray}
The first term on the right hand of Eq. \ref{eq:decompose_d} is the electron translation factor (ETF) term\cite{athavale2023} defined by:
\begin{eqnarray}
\label{eq:detf}
    d_{IJ}^{A \alpha,ETF} &=& \sum_{\mu \nu} \frac{1}{2} \left( \left<\chi_{\mu} \middle| \frac{\partial}{\partial R_{A \alpha}} \chi_{\nu} \right> - \left<\chi_{\nu} \middle| \frac{\partial}{\partial R_{A \alpha}} \chi_{\mu} \right> \right) D_{\mu \nu}
    %&=&\frac{1}{2i\hbar} \text{Tr}\left(D \cdot p^{\alpha} ( \delta_{BA} + \delta_{CA})\right)
\end{eqnarray}
where $D_{\mu \nu}$ is the one-particle density matrix.
(For a standard Hartree-Fock or Kohn-Sham DFT calculation,  with molecular orbitals $C_{\mu i}$, one defines $D_{\mu\nu} = \sum_{i}^{n_{orb}} C_{\mu i}C_{\nu i}^{*}$.)
In this separation, it is crucial to emphasize that
whereas $\tilde{\bd}_{IJ}$ diverges near a conical intersections, $\bd_{IJ}^{ETF}$ is always small even near a conical intersection; more precisely, whereas the size of the former depends on the inverse of the energy difference between electronic states, the size of the latter does not. Moreover, because the atomic orbital basis $\left\{ \chi_{\mu} \right\}$ is always chosen to be real-valued, there is no phase problem in isolating the term $\bd_{IJ}^{ETF}$. Third, differentiating such a term is computationally trivial.
For all of these reasons, it is clear that if one were to approximate $\bd$ by $\bd^{ETF}$ in Eq. \ref{eq:shenvi}, the resulting phase-space Hamiltonian would appear to be much more stable and tractable.

Now, in Sec. \ref{sec:Outline} above, we discussed briefly the fact that standard BO dynamics along the surface $E_K(\bR)$ (generated by Eq. \ref{eq:standard}) fails to conserve either linear or angular momentum. In brief, because BO dynamics propagate nuclear dynamics along translationally and rotationally invariant potential energy surfaces, these dynamics conserve the {\em nuclear} linear and angular momentum. At the same time, however, these algorithms ignore the electronic linear and angular momentum and do not conserve the {\em total} linear or angular momentum (provided the electronic observables are nonzero). Previously, in Ref. \citenum{coraline24gamma} and \citenum{tian24gamma}, we showed that in order to conserve the total linear and angular momentum along a phase-space adiabat, one can construct a Hamiltonian of the following form:

\begin{eqnarray}
    \hH_{PS} &\equiv& \sum_{A\alpha} \frac{P_{A\alpha}^2}{2 M_A} + \hH_{el} - \sum_{\mu \nu A\alpha} i \hbar \frac{P_{A\alpha}}{M_A} \cdot \bar{\Gamma}^{A\alpha}_{\mu \nu} a_{\mu}^{\dagger} a_{\nu} \label{eq:Hps1} \\
    \bar{\bGamma}_{\mu \nu}^A &\equiv& \sum_{
    \lambda  \sigma}  S^{-1}_{\mu \lambda}  \bGamma^{A}_{\lambda \sigma} S^{-1}_{\sigma\nu } \label{eq:Hps2}
\end{eqnarray}
provided the $\bGamma$ operators in Eq. \ref{eq:Hps2} are antisymmetric and satisfy:
 \begin{align}
 \label{eq:Gamma_uv1}
     -i\hbar\sum_{A}\Gamma^{A\alpha}_{\mu\nu} + p^{\alpha}_{\mu\nu} &= 0  \\
      \label{eq:Gamma_uv2}
\sum_{B}\nabla_{B\beta}\Gamma^{A\alpha}_{\mu\nu} &= 0\\
     \label{eq:Gamma_uv3}
    -i\hbar\sum_{A\beta\gamma}\epsilon_{\alpha\beta\gamma}{X}_{A\beta} \Gamma^{A\gamma}_{\mu\nu} + l^{\alpha}_{\mu\nu} + s^{\alpha}_{\mu\nu} &= 0
    \\
     \sum_{B\beta\eta} \epsilon_{\alpha\beta\eta} X_{B\beta} \bra{\mu} \frac{\partial \hat{\Gamma}^{A\gamma}}{\partial X_{B\eta}}\ket{\nu} -\frac{i}{\hbar}
        \bra{\mu} [\hat{\Gamma}^{A\gamma},\hat{L}^{\alpha}_{e} ]\ket{\nu} + \sum_{\eta}\epsilon_{\alpha\gamma\eta} \Gamma^{A\eta}_{\mu\nu} &= 0 \label{eq:Gamma_uv4}
 \end{align} 
where
$p^{\alpha}_{\mu\nu}$ is the linear momentum  in the atomic orbital basis,
and $l^{\alpha}_{\mu\nu}$ and $s^{\alpha}_{\mu\nu}$ are the angular momentum and spin matrices in the atomic orbital basis (respectively). Several comments are now in order. 

\begin{enumerate}
    \item First, note that in Eq. \ref{eq:Hps2}, the $S$ matrix is the overlap matrix. If one had access to a localized, atom-centered orthonormal basis, one could ignore this term entirely (as $S$ would be the identity). In conventional quantum chemistry calculations, however, $S$ is not the identity, and including the $S^{-1}$ factors in Eq. \ref{eq:Hps2} is one means to ensure that the final energy should involve contractions of the $\bGamma_{\mu \nu}$ matrices. Indeed, if one calculates the expectation value for the operator  $\sum_{\mu \nu} \bar{\bGamma}^{A}_{\mu \nu} a_{\mu}^{\dagger} a_{\nu}$ in a state $\ket{\Psi}$, one finds:
\begin{eqnarray}
    \left<  \Psi \middle| \sum_{\mu \nu} \bar{\bGamma}^{A}_{\mu \nu} a_{\mu}^{\dagger} a_{\nu} \middle| \Psi \right> =     
    \sum_{\mu \nu \lambda \sigma} \bGamma^{A}_{\lambda \sigma} S^{-1}_{\sigma \mu} \left<  \Psi \middle|  a_{\mu}^{\dagger} a_{\nu} \middle| \Psi \right> S^{-1}_{\nu \lambda} = \sum_{\lambda \sigma} \bGamma^{A}_{\lambda \sigma} D_{\sigma\lambda }
\end{eqnarray}
where $D_{\sigma\lambda }$ is defined in Eq. \ref{eq:detf}.

\begin{comment}

where $D_{\lambda \sigma}$ is the usual definition of a one-electron density matrix in a non-orthogonal atomic basis. For instance, if the state $\ket{\Psi}$ is a Slater determinant with orbital coefficient $C_{\mu i}$, one finds the usual definition

\begin{eqnarray}
    D_{\lambda \sigma} = \sum_i C_{\lambda i} C_{\sigma i}
\end{eqnarray}

\end{comment}

\item Second, the intuition behind Eqs. \ref{eq:Gamma_uv1}-\ref{eq:Gamma_uv4} is as follows. Eqs. \ref{eq:Gamma_uv1} and \ref{eq:Gamma_uv3} are phase conventions that mathematically encapsulate our requirement that electronic wavefunctions be functions of the electronic coordinates relative to the nuclear coordinates. Thus, Eq. \ref{eq:Gamma_uv1} stipulates that under translation of the nuclei, the electronic wavefunction will also translate. Eq. \ref{eq:Gamma_uv3} stipulates that under rotation of the nuclei, the entire electronic wavefunction will also rotate.
Next, Eqs. \ref{eq:Gamma_uv2} is a mathematical statement that the $\bGamma$ couplings should effectively be invariant to translations of the molecule, $\Gamma^{A \alpha}_{\mu \nu}(\bR_0 + \Delta) = \Gamma^{A \alpha}_{\mu \nu}(\bR_0)$.  Eqs. \ref{eq:Gamma_uv1}-\ref{eq:Gamma_uv3} are closely related to the transformation properties of the basis states and derivative couplings under translations and rotations as discussed in Refs. \citenum{LittlejohnRawlinsonSubotnik23,LittlejohnRawlinsonSubotnik24}, the main difference being that those papers treat multielectron basis functions rather than single particle orbitals.  For example, Eq. \ref{eq:Gamma_uv1} and \ref{eq:Gamma_uv2} are closely related to Eqs.~(172) and (179) of Ref. \citenum{LittlejohnRawlinsonSubotnik24} and Eq. \ref{eq:Gamma_uv3} is related to Eq.~(107) of Ref. \citenum{LittlejohnRawlinsonSubotnik23}.

In Eq. \ref{eq:Gamma_uv4}, the operator $\hat{\Gamma}^{A\gamma}$ can be any operator that satisfies
$\left< \mu \middle| \hat{\Gamma}^{A\gamma} \middle| \nu \right> = {\Gamma}_{\mu \nu}^{A\gamma} $.
Note that the form of Eq. \ref{eq:Gamma_uv4} is slightly involved because within usual quantum chemistry calculations with atomic orbital basis functions, one does not reorient a given atomic orbital upon rotation. For instance, a $p_x$ orbital is always in the $x$ direction of the lab frame no matter the orientation of the molecule.
Nevertheless, despite this annoyance, the meaning of Eq. \ref{eq:Gamma_uv4} is clear: the matrix elements must be invariant to rotations. In other words, if 
$\mathsf{U}$ is a rotation matrix, Eq. \ref{eq:Gamma_uv4} dictates that
 $\Gamma^{A \alpha}_{\mathsf{U}\mu \mathsf{U} \nu} (\mathsf{U}\bR_0) = \sum_{\beta} \mathsf{U}_{\alpha \beta} \Gamma^{A \beta}_{\mu \nu}(\bR_0)$.

\item Third, if one were to fix $\bGamma^{A}$ as
\begin{eqnarray}
 \bGamma'^{A}_{\mu \nu} = \frac{1}{2} \left( \left<\chi_{\mu} \middle| \frac{\partial}{\partial R_{A \alpha}} \chi_{\nu} \right> - \left<\chi_{\nu} \middle| \frac{\partial}{\partial R_{A \alpha}} \chi_{\mu} \right> \right)
\end{eqnarray}
in the spirit of Eq. \ref{eq:detf}, 
one can show that this definition of $\bGamma^A$ satisfies the translational requirements in Eq. \ref{eq:Gamma_uv1} and \ref{eq:Gamma_uv2}. However, such a definition does not satisfy Eqs. \ref{eq:Gamma_uv3} - \ref{eq:Gamma_uv4} for rotations. However, as demonstrated in Ref. \citenum{coraline24gamma} and recapitulated in Appendix \ref{appendix:gammaredefine}, one can satisfy all of the necessary requirements by fixing
\begin{eqnarray}
\label{eq:fullgamma}
    \bGamma^A_{\mu \nu}  = \bGamma'^A_{\mu \nu}  +  \bGamma''^A_{\mu \nu}  
\end{eqnarray}
where we refer to $\bGamma''^A_{\mu \nu}$ as an electron rotational factor (ERF).
It should be noted that $\bGamma^A$ in Eq. \ref{eq:fullgamma} does not diverge (just like $\bm{d_{IJ}^{A,ETF}}$), and therefore this ansatz remains a suitable candidate for an electronic-nuclear momentum coupling term. 
\end{enumerate}

%\noindent From these basic set of conditions, we derived a one-electron operator $\bm \Gamma_{\mu \nu}$ using electron-translation and electron-rotation factors. By including the first order $\bm \Gamma \cdot \bm P$ term in our electronic Hamiltonian, we are guaranteed to not only conserve total linear and angular momenta, but critically, allow exchange between the nuclear and electronic momenta components, thereby coupling them. Like the full derivative coupling, this new operator was shown to satisfy the phase and translational/rotational invariance conditions of Eq. \ref{eq:Gamma_uv4}, but, unlike the full derivative coupling, this operator is computationally inexpensive, and does not diverge near surface crossings.\cite{izmaylov:2016:jpc_dboc_pssh} This term therefore not only gives us a way to capture a significant portion of the coupling between nuclear and electronic motion, but also gives us a cheap and stable way to obtain it. 

 Thus, at the end of the day, using a standard electronic Hamiltonian with a one-electron operator $\hat{h}$ and a two electron operator $\hat{\pi}$, the most naive phase-space Hamiltonian (as encoded by the Hamiltonian in Eq. \ref{eq:Hps1} above) postulates that the molecular energy for a system at positions and momenta $(\bR,\bP)$ is of the form:
\begin{equation}
\label{eq:eps_unstable}
E_{\text{PS}}(\bR,\bP) = \sum_{A\alpha}\frac{P_{A\alpha}^2}{2M_A} + \sum_{\mu\nu}D_{\nu\mu} \Big(h_{\mu\nu} - i\hbar \sum_{A\alpha} \frac{ P_{A\alpha}\Gamma^{A\alpha}_{\mu\nu}}{M_{A}}\Big) + \sum_{\mu\nu\lambda\sigma}G_{\nu\mu\sigma\lambda}\pi_{\mu\nu\lambda\sigma}
\end{equation}
Here $G_{\nu\mu\sigma\lambda}$ represents the two-electron density matrix in an atomic orbital basis.

\subsection{Including a Second-Derivative Coupling}\label{sec:gamma2}

Unfortunately, we have found that the the energy expression in Eq. \ref{eq:eps_unstable} is not stable as far as generating VCD spectra, especially for large basis sets. After some bench-marking, our tentative hypothesis is that problems arise if the kinetic momentum 
$$\sum_{A\alpha}\frac{P_{A\alpha}^2}{2M_A} - i\hbar \sum_{\mu\nu} D_{\nu\mu} \sum_{A\alpha} \frac{ P_{A\alpha}\Gamma^{A\alpha}_{\mu\nu}}{M_{A}} $$
is not positive definite (as it must be).
 To that end, given the form of the exact Hamiltonian in Eq. \ref{eq:exactBOH} in an adiabatic representation, we posit that a second-order term is appropriate. 
 
 At this point, let us consider the simplest possible system: a single electron interacting with a host of nuclei. In such a situation, second-quantization becomes trivial and equivalent to first quantization, and we hypothesize that one can write down a meaningful analogue of Eq. \ref{eq:exactBOH} in second-quantization as follows:

\begin{eqnarray}
\label{eq:temp1}
\hH_{PS} &=& \sum_{\mu \nu} \bar{T}^{nuc}_{\mu \nu}    a_{\mu}^{\dagger} a_{\nu} + \hH_{el} \\
\label{eq:temp2}
\bar{T}^{nuc}_{\mu \nu} &=& \sum_{\lambda \sigma} S^{-1}_{\mu \lambda} T^{nuc}_{\lambda \sigma} S^{-1}_{\sigma \nu}
\\
    T^{nuc}_{\mu \nu} &\stackrel{?}{=}& \sum_{A \lambda \sigma} \frac{1}{2M_A} (\bP_A \delta_{\mu \lambda} - i\hbar \tilde{\Gamma}^A_{\mu \lambda}) S_{\lambda \sigma} (\bP_A \delta_{\sigma \nu} - i \hbar \tilde{\Gamma}^A_{\sigma \nu})
    \label{eq:maybe}
\end{eqnarray}
where we fix $\tilde{\bGamma}$ to be the solution to:
\begin{eqnarray}
\label{eq:Gtildedef}
\frac{1}{2} \sum_{\lambda} \bm{\tilde{\Gamma}^A_{\mu \lambda}} S_{\lambda \nu} + S_{\mu \lambda} \bm{\tilde{\Gamma}^A_{\lambda \nu}} = \bm{\Gamma^A_{\mu \nu}}
\end{eqnarray}

Note that, just as in Eq. \ref{eq:shenvi}, Eq. \ref{eq:maybe} expresses the kinetic energy in product form and posits that the function $\bP_A - \tilde{\bGamma}^A_{\mu \lambda}$ represents a kinetic momentum operator; intuitively, the $\tilde{\bGamma}^A_{\mu \lambda}$ captures how 
electrons are being dragged by a nucleus whenever that nucleus moves -- because each nucleus carries a basis function along for the ride.
Unfortunately, however, this operator only makes sense for a one electron problem. After all, if one inserts Eq. \ref{eq:maybe} into Eq. \ref{eq:temp2} and then into Eq. \ref{eq:temp1}, one will find that $\hat{H}_{PS}$ contains a term proportional to $\sum_{A} \frac{\bP_A \cdot \bP_A}{2M_A}\sum_{\mu \nu} S^{-1}_{\mu \nu} a_{\mu}^{\dagger} a_{\nu}$, and it is easy to show that for any wavefunction, $\left<S^{-1}_{\mu \nu} a_{\mu}^{\dagger} a_{\nu}\right> = N_e$, where $N_e$ is the number of electrons.
Thus, Eqs. \ref{eq:temp1}-\ref{eq:maybe} cannot be a meaningful Hamiltonian for a many-electron system where $N_e \ne 1$.
Nevertheless, we can now identify one plausible many-body Hamiltonian that (i) reduces to product form for the case of a single electron, (ii) maintains angular and linear momentum conservation and (iii) has a positive definite kinetic energy (See Appendix \ref{appendix:posdef}). Namely, we can make the ansatz that a reasonable semiclassical phase-space Hamiltonian is:
\begin{eqnarray}
\label{eq:finalH1}
\hH_{PS} &=&  \hH_{el} + \frac{\bP_A \cdot \bP_A}{2M_A} -i \hbar  \sum_{A \alpha \mu \nu } \frac{P_{A \alpha} \cdot \bar{\Gamma}^{A\alpha}_{\mu \nu} }{M_A}   a_{\mu}^{\dagger} a_{\nu}
+ \sum_{A \alpha \mu \nu } \frac{\bar{\zeta^{A\alpha}_{\mu \nu}} }{M_A}   a_{\mu}^{\dagger} a_{\nu}\\
\label{eq:finalH2}
\bar{\bGamma}^{A}_{\mu \nu} &=& \sum_{\lambda \sigma} S^{-1}_{\mu \lambda}  \bGamma^{A}_{\lambda \sigma}   S^{-1}_{\sigma \nu} \\
\label{eq:finalH3}
\bar{\zeta^{A\alpha}_{\mu \nu}} &=& -\hbar^2 
\sum_{\lambda \sigma \kappa \eta} S^{-1}_{\mu \lambda}  \tilde{\Gamma}^{A\alpha}_{\lambda \kappa} S_{\kappa \eta} \tilde{\Gamma}^{A\alpha}_{\eta \sigma}  S^{-1}_{\sigma \nu}
\end{eqnarray}
where $\tilde{\bGamma}_{\mu \nu}$ is defined in Eq. \ref{eq:Gtildedef} and we show how to solve this equation in 
In Appendix \ref{appendix:posdef}. 
Note that our explicit form for $\bGamma$ is defined in Appendix \ref{appendix:gammaredefine} (following Ref. \citenum{coraline24gamma}). Eqs. \ref{eq:finalH1}-\ref{eq:finalH3} represent the final form of our proposed phase-space Hamiltonian. Just as in Eq. \ref{eq:eps_unstable} above, one can easily write down the final energy for this Hamiltonian:

\begin{equation}
\label{eq:eps_stable}
E_{\text{PS}}(\bR,\bP) = \sum_{A\alpha}\frac{P_{A\alpha}^2}{2M_A} + \sum_{\mu\nu}D_{\nu\mu} \Bigg(h_{\mu\nu} - i\hbar \sum_{A\alpha} \frac{ P_{A\alpha}\Gamma^{A\alpha}_{\mu\nu}}{M_{A}}  - \sum_{A\alpha\lambda\sigma}  \hbar^2 D_{\nu\mu}\frac{\tilde{\Gamma}^{A\alpha}_{\mu\sigma} S_{\sigma\lambda} \tilde{\Gamma}^{A\alpha}_{\lambda\nu}}{2M_{A}}\Bigg) + \sum_{\mu\nu\lambda\sigma} G_{\nu\mu\sigma\lambda}\pi_{\mu\nu\lambda\sigma}
\end{equation}

Note that, according to Ref. \citenum{coraline24gamma}, $\bGamma$ acts like an effective vector potential (which generates an effective magnetic field), so that the canonical momentum ($\bP$) is not equal to the kinetic momentum except at ($\bm{\Pi} = \bm{P} = \bm{0}$), and even then, the derivatives $\pp{}{\Pi_{A\alpha}} \neq \pp{}{P_{A\alpha}}$. In particular, 

\begin{equation}
\begin{aligned}
\label{eq:canonicalkineticp}
\dot{X}_{A\alpha} =& \pp{E_{PS}}{P_{A\alpha}} = \frac{P_{A\alpha}}{M_A} - i\hbar \text{Tr}\left(\frac{\Gamma^{A\alpha} D}{M_A}\right) \\
\ne &  \frac{P_{A\alpha}}{M_A}
\end{aligned}
\end{equation}

\subsection{The Harmonic Approximation and Normal Modes for Phase-Space Electronic Hamiltonians} \label{sec:PSModes}

The basis of eigenvectors from the phase-space electronic Hamiltonian in Eq. \ref{eq:finalH2}, 
\begin{eqnarray}
\hH_{PS}(\bR, \bP) \ket{\Phi^{PS}_J} =      E^{PS}_J(\bR, \bP) \ket{\Phi^{PS}_J}
\end{eqnarray}
gives a new basis for expansion of the total nuclear-electronic wavefunction, 
\begin{eqnarray}
\ket{\Psi_{tot}(t)} =  \int d\bR \int d\bP \sum_J \Omega_J(\bR,\bP,t) \ket{\bR,\bP}\ket{\Phi^{PS}_J;\bR,\bP}
\end{eqnarray}
where $\ket{\bR,\bP}$ is a coherent nuclear state. To zeroth order, however, the first and most important step is usually to make a one-state approximation and extract vibrational energies. And in that vein, one usually begins by expanding the potential surfaces to second order and generating normal modes and vibrational frequencies.

\subsubsection{Approach to Normal Modes in a Born-Oppenheimer Representation}
Before addressing how to construct phase-space normal modes, let us review for the usual notation for 
a Born-Oppenheimer expansion. One defines $V(\bR) = E_G(\bR)$ (the lowest eigenvalue of $\hat{H}_{el}$ in Eq. \ref{eq:standard}) and expands to second order in displacements from equilibrium:
\begin{eqnarray}
V(\bR) = V(\bm{R_{eq}}) + \frac{1}{2}\sum_{A \alpha, B \beta} K_{A \alpha, B\beta} (R_{A \alpha} - R^{eq}_{A \alpha})(R_{B \beta} - R^{eq}_{B \beta})
\end{eqnarray}
If now one regards the quantities $\left\{ R_{A \alpha} \right\}$ as operators (rather than scalars), i.e. one quantizes these operators, the vibrational Hamiltonian is then defined as:
\begin{eqnarray}
\hH_{vib} &=& \frac{1}{2M} \bhP \cdot \bhP + V(\bhR) \\
&=&  \sum_{A \alpha} \frac{1}{2M_A}\hP_{A\alpha}^2 + V(\bR_{eq}) + \frac{1}{2} \sum_{A \alpha, B \beta} K_{A \alpha, B\beta} (\hR_{A \alpha} - R^{eq}_{A \alpha})(\hR_{B \beta} - R^{eq}_{B \beta})
\end{eqnarray}
where $K_{A \alpha, B\beta} = \frac{\partial^2 V}{\partial R_{A\alpha}\partial R_{B\beta}}$ (i.e. $\bK^{BO}= \frac{\partial^2 V}{\partial \bR \partial \bR}$ in matrix form)
is the Hessian matrix. Normal modes are generated by shifting to relative coordinates (setting $\bR^{eq} = 0$), ignoring equilibrium energy (C-number), and converting to mass-weighted coordinates given by
\begin{eqnarray}
    \bhR &=& \bhR'' M_{BO}^{-1/2} \\
    \bhP &=& \bhP'' M_{BO}^{1/2} 
\end{eqnarray}
where we define $M_{BO}$ as the diagonal matrix of nuclear masses, which eliminates the masses in the kinetic energy. This leaves
\begin{eqnarray}\label{eq:diagPS}
\hH_{vib} &=& \frac{1}{2} \bhP''^{\dagger}\bhP'' + \frac{1}{2} \bhR''^{\dagger} \bK'' \bhR''
\end{eqnarray}
where $\bK'' = M_{BO}^{-1/2} \bK M_{BO}^{-1/2}$.
Finally, one diagonalizes the resulting Hamiltonian $\bK''$ and the result is a series of uncoupled mass weighted normal-mode oscillators.

\subsubsection{Approach to Normal Modes in a Phase-Space Representation}
The procedure above can be largely replicated in a phase-space picture.
One expands the lowest eigenvalue of our phase space Hamiltonian $E_{PS}(\bR,\bP)$ in Eq. \ref{eq:eps_stable} to second order (in both $\bR$ and $\bP$), quantizes them, and then one finds:

\begin{equation}
\begin{aligned}
\label{eq:tot_nuc_hess} 
\hH_{vib} 
&= E_{PS}(\bR^{eq},\bP^{eq}) + \frac{1}{2}  \sum_{A \alpha, B \beta} W_{A \alpha, B\beta} (\hP_{A \alpha} - P^{eq}_{A \alpha})(\hP_{B \beta} - P^{eq}_{B \beta})   \\
& +\frac{1}{2} \sum_{A \alpha, B \beta} K_{A \alpha, B\beta} (\hR_{A \alpha} - R^{eq}_{A \alpha})(\hR_{B \beta} - R^{eq}_{B \beta})  \\
&+  \frac{1}{2} \sum_{A \alpha, B \beta} Y_{A \alpha, B\beta} \left( (\hR_{A \alpha} - R^{eq}_{A \alpha})(\hP_{B \beta} - P^{eq}_{B \beta}) +
(\hP_{A \alpha} - P^{eq}_{A \alpha})(\hR_{B \beta} - R^{eq}_{B \beta}) \right)
%\hH_{vib}(R, P) &= R^{+}\mathcal{H}^{RR} R + R^\dagger\mathcal{H}^{RP} P + P^{+}\mathcal{H}^{PR} R+ P^{+}\mathcal{H}^{PP} P\\
%&= R^{+}\mathcal{H}^{RR} R + P^{+}\mathcal{H}^{PP} P
\end{aligned}
\end{equation}
where $W_{A\alpha,B\beta} = \left(\frac{E_{PS}}{\partial P_{A\alpha}\partial P_{B\beta}}\right)_{eq}$, $K_{A\alpha,B\beta} = \left(\frac{E_{PS}}{\partial R_{A\alpha}\partial R_{B\beta}}\right)_{eq}$, and $Y_{A\alpha,B\beta} = \left(\frac{E_{PS}}{\partial P_{A\alpha}\partial R_{B\beta}}\right)_{eq}$.
See Appendix \ref{appendix:DiagonalHess} for explicit forms of $W_{A\alpha,B\beta}, K_{A\alpha,B\beta}, \mbox{ and } Y_{A\alpha,B\beta}$. Note that, for this paper, we do not invoke spin degrees of freedom. This assumption implies that $\bm{Y}=0$ and $\bP_{eq} = 0$ (see Appendix \ref{appendix:DiagonalHess}). Shifting to relative positional coordinates and ignoring equilibrium energy, the total vibrational Hamiltonian is of the form:

\begin{eqnarray}
\label{eq:Hvibtmp}
\hH_{vib} &=& \frac{1}{2} \bhP^{\dagger}\bm{W}\bhP + \frac{1}{2} \bhR^{\dagger} \bK \bhR
\end{eqnarray}

Unfortunately, unlike in the BO representation, $\bW$ is not diagonal by construction. Thus, in order to find the normal modes which diagonalize our separable phase-space Hamiltonian, we first need to find the unitary transformation which diagonalizes $\bW = U \Delta U^\dagger$ (where $\Delta \approx M_{BO}^{-1}$ for small perturbations to the uncoupled momentum Hessian). Taking the coordinate transformation $\bR = U \bR'$, $\bP = U \bP'$, Eq. \ref{eq:Hvibtmp} can then be written as

\begin{eqnarray}
\label{eq:before:mass}
\hH_{vib} &=& \frac{1}{2} \bhP'^{\dagger}\Delta\bhP' + \frac{1}{2} \bhR'^{\dagger} U^{\dagger} \bK U\bhR'
\end{eqnarray}

%\begin{equation}
%H_{nuc}(R',P') = R'^\dagger U^\dagger H^{R} U R' + P'^\dagger \tilde{M}^{-1} P'
%\end{equation}

\noindent Eq. \ref{eq:before:mass} is diagonal in momenta but with new effective masses. Thus, if we again change to mass-weighted coordinates $\bR' = \Delta^{1/2}\bR''$, $\bP' = \Delta^{-1/2}\bP''$, we find
%\noindent Then suppose that we change coordinates once again to mass-weighted coordinates $X' = \tilde{M}^{-1/2}\rho$, $P' = \tilde{M}^{1/2}\pi$, we then find
\begin{eqnarray}
\hH_{vib} &=& \frac{1}{2} \bhP''^{\dagger}\bhP'' + \frac{1}{2} \bhR''^{\dagger} \Delta^{1/2} U^{\dagger} \bK U \Delta^{1/2} \bhR''\\
&=& \frac{1}{2} \bhP''^{\dagger}\bhP'' + \frac{1}{2} \bhR''^{\dagger} \bK'' \bhR''
\end{eqnarray}
which is of the same form as the standard quadratic expansion in the BO representation (Eq. \ref{eq:diagPS}). If $\bm{k''}$ are the eigenvectors of $\bK''$, 
%which represent the mass-weighted nuclear normal modes in a non-BO representation, 
the nuclear normal modes in our original Cartesian coordinates are then simply
\begin{equation}
\bm{k} = U \Delta^{1/2} \bm{k}''
\end{equation}

After diagonalization, we write the final form for the Hamiltonian in normal coordinates as the sum of independent harmonic oscillators:

\begin{align}
\label{eq:massmodesQP}
    \hat{H}_{vib} = \frac{1}{2} \bm{\hat{\mathcal{P}}}_k^2  + \sum_k \frac{1}{2}w_k^2 \bm{\hat{Q}}_k^2 
\end{align}

\begin{comment}

\begin{equation}
\begin{split}
H_{nuc}(\rho, \pi) &= \rho^+ \tilde{M}^{-1/2}U^+ \mathcal{H}^{RR} U \tilde{M}^{-1/2} \rho + \pi^+ \tilde{M}^{1/2} \tilde{M}^{-1} \tilde{M}^{1/2} \pi \\
&= \rho^+ \tilde{\mathcal{H}}^{RR} \rho + I\\
H_{nuc}(\rho)  &= \rho^+ \mathcal{H}^{RR} \rho\\
\end{split}
\end{equation}

\noindent Where $\pi$ is some normalized vector $\therefore \pi \pi^{+} = I$. 

If $\rho$ are eigenvectors of $\tilde{\mathcal{H}}^{RR}$, the nuclear normal modes in our original coordinates are then 

\begin{equation}
R = U \tilde{M}^{-1/2} \rho
\end{equation}
\end{comment}

\subsection{Rotatory Strength in Phase Space Framework}\label{sec:rotstr}

\noindent If the interaction between an electromagnetic field with a molecular system is given by $\hH_{int} = -\bm{\mhb} \cdot \bm{E}_{ext} -\mmhb \cdot \bm{B}_{ext}$, then when averaged over all incoming light directions, the differential absorbance to circularly polarized light of chiral vibrations is given by the rotatory strength\cite{power}:

\begin{equation}
\label{eq:rot_str}
\begin{aligned}
    \mathcal{R} = Im[\bra{\Psi_{Gg}} \bm{\hat{\mathcal{m}}} \ket{\Psi_{Ge}}\cdot \bra{\Psi_{Ge}} \bm{\hat{\mu}} \ket{\Psi_{Gg}}]
\end{aligned} 
\end{equation}

\noindent The magnetic moment $\bm{\mathcal{m}}$ and electric moment $\bm{\mu}$ each contains an electronic and nuclear component. Ignoring spin contributions, these operators are given as 

\begin{eqnarray}
    \label{eq:medefinition}
    \mmhb^e &=& \frac{-e}{2m_e} \bm{\hat{r}} \times \bm{\hat{p}} =  \frac{-e}{2m_e} \bm{\hat{L}^e} \\
    \label{eq:mndefinition}
    \hat{\bm{\mathcal{m}}}^n  &=& \sum_{A} \frac{Z_{A}e}{M_A 2} \bhR_{A} \times \bhP_{A}\\
    \label{eq:muedefinition}
    \mhb^e &=& - e \bm{\hat{r}} \\
    \label{eq:mundefinition}
    \hat{\bm{\mu}}^n &=& \sum_{A} Z_{A} e \bhR_{A}
\end{eqnarray}

\noindent Note that in Eqs. \ref{eq:medefinition} - \ref{eq:mundefinition} we use the standard $\bm{r},\bm{p},\bR,\bP$ coordinates to express the magnetic and electric moments.
These expressions ignore the fact that formally these operators are dressed under a diabatic-to-adiabatic transformation and, e.g., convert a canonical momentum $\bP$ to a kinetic momentum $\bm{\Pi}$. Nevertheless, one expects these dressings to be small and so they are ignored at this level of treatment.
%This might appear contradictory insofar as the fact that we have made an adiabatic transformation and one can question, e.g., why we do not use the kinetic momentum rather than the canonical momentum.  That being said, this approximation is fairly standard in the literature and is simply a reflection of the fact that
For instance, following Eq. \ref{eq:canonicalkineticp}, it follows that
%the canonical momentum is used rather than a properly dressed momentum; for instance, one might assume that one should use the kinetic momentum instead of the canonical momentum for the nuclei. Nevertheless, (which one might assume should be the quantity which $\bm{\mathcal{m}}$ is formally associated with. While these quantities are the identical for the electronic momenta, the nuclear canonical and kinetic momenta are not. Fortunately these quantities are connected by the equation
\begin{equation}
\left(\pp{\bm{\Pi}}{\bm{P}}\right)_R = 1 - i\hbar \text{Tr}(D^{[P]}\cdot \bm{\Gamma})_{R} \approx 1
\end{equation}
where the subscript $R$ indicates that nuclear coordinates are kept constant, $D^{[P]}$ is the momentum derivative of the density matrix, and we assume $\Gamma$ is small. With this caveat in mind, let us evaluate $\bm{\mu}$. For a transition between product nuclear-electronic wavefunctions (in the spirit of the BO approximation), we can evaluate 
\begin{equation}
\label{eq:BO_expand}
\begin{aligned}
    \bra{\Psi_{Gg}} \bm{\hat{\mu}} \ket{\Psi_{Ge}} &=  \bra{\Omega_{Gg}} \otimes \bra{\Phi_{G}} \mhb^e \ket{\Phi_{G}} \otimes \ket{\Omega_{Ge}} + \bra{\Omega_{Gg}} \bm{\hat{\mu}}^n \ket{\Omega_{Ge}}\\
    & \equiv \bra{\Omega_{Gg}}  \hat{\bm{\mu}}_{G}  \ket{\Omega_{Ge}}
\end{aligned}
\end{equation}
%where $\bmu_G(\bR) = \left<\Phi_G\middle | \mhb^e \middle | \Phi_G \right> + \bm{\mu}^n$.
If one neglects the dependence of $\bmu_G$ on $\bR$ (evaluates $\bmu_G$ at $\bm{R_{eq}}$), then Eq. \ref{eq:BO_expand} vanishes by the orthogonality of $\Omega_{Gg}$ and $\Omega_{Ge}$. In order to obtain a non-zero rotatory strength, one must expand in nuclear coordinates around $\bR = \bR_{eq}$. Doing so and invoking the harmonic approximation yields the following expression for the transition electric dipole moment.

\begin{equation}
\begin{aligned}
\label{eq:mu_pet}
    \bra{\Omega_{Ge}} \otimes \bra{\Phi_{G}} \bm{\hat{\mu}}  \ket{\Phi_{G}} \otimes \ket{\Omega_{Gg}} &= \left< \Omega_{Ge} \middle|  (\bm{\mu_{G}})_{eq}  + \sum_{A\alpha} \left( \frac{\partial \bm{\mu_{G}}}{\partial R_{A \alpha}}  \right)_{eq} \cdot (\hat{R}_{A \alpha}-R^{eq}_{A \alpha}) \middle| \Omega_{Gg} \right>\\
    & = \sum_{A\alpha} \left( \frac{\partial \bm{\mu_{G}}}{\partial R_{A \alpha}}  \right)_{eq}\cdot  \bra{\Omega_{Ge}} (\hat{R}_{A \alpha}-R^{eq}_{A \alpha}) \ket{\Omega_{Gg}}\\
    &= \sum_k  \sum_{A\alpha }\left( \frac{\partial \bm{\mu_{G}}}{\partial R_{A \alpha}}  \right)_{eq}\left(\frac{\hbar}{2M_A\omega_k}\right)^{1/2} \mathcal{S}_{A \alpha,k}
\end{aligned}
\end{equation}
Here the element $\mathcal{S}_{A \alpha,k}$ quantifies the displacement of nucleus $A$ in direction $\alpha$ for the $k^{th}$ vibrational mode and is defined as 
\begin{align}
\mathcal{S}_{A \alpha,k}= \frac{\partial R_{A\alpha}}{\partial Q_k} = \frac{\partial \dot{R}_{A\alpha}}{\partial \mathcal{P}_k} = \frac{1}{M_A} \frac{\partial P_{A\alpha}}{\partial \mathcal{P}_k}
\end{align}
where $Q_k$ and $\mathcal{P}_k$ are the mass-weighted position modes and mass-weighted momentum modes from Eq. \ref{eq:massmodesQP}. This completes our treatment of $\pp{\bmu}{\bR}$. 
%due to the symmetry of these operators and of the ground state wavefunction, $\frac{\partial^n \bm{\mu}}{(\partial P)^n} = 0 \ \forall \ n$ (given that ${\bf{P}}^{eq} = 0$)

Next, we turn to $\bra{\Psi_{Gg}} \hat{\bm{\mathcal{m}} }\ket{\Psi_{Ge}}$. At the equilibrium geometry ($\bR=\bR_{eq}$) and equilibrium momentum ($\bP = 0$), $\bra{\Phi_{G}} \bm{\hat{\mathcal{m}}}^e(\bR_{eq},0) \ket{\Phi_{G}} = 0$ since $\bm{\hat{\mathcal{m}}}^e$ is a purely imaginary Hermitian operator and $\Phi_{G}$ can be chosen real for a time-reversible ground state. Furthermore, $\left<\bm{\hat{\mathcal{m}}}^n\right>$ also clearly vanishes if $\bP = 0$. We conclude that $\bm{m}$ vanishes for all nuclear geometries unless $\bP \ne 0$, and thus expand $\bm{m}$ in the canonical momentum $\bP$:

%which is related to the kinetic momentum by $\Pi_{A\alpha} = P_{A\alpha} - i\langle \Gamma_{A\alpha} \rangle$. For real wave-functions, $\langle \Gamma_{A\alpha} \rangle = 0$, thus we simply expand in P. 

\begin{equation}
\label{eq:L_pet}
\begin{aligned}
    \bra{\Omega_{Gg}} \bra{\Phi_{G}} \bm{\hat{\mathcal{m}}} \ket{\Phi_{G}} \ket{\Omega_{Ge}} &= \left< \Omega_{Gg} \middle| (\bm{\mathcal{m}_{G}})_{eq}  + \sum_{A\alpha} \left( \frac{\partial \bm{\mathcal{m}_{G}}}{\partial P_{A \alpha}}\right)_{eq} \cdot \hat{P}_{A \alpha} \middle| \Omega_{Ge} \right>\\
    & = \sum_{A\alpha} \left( \frac{\partial \bm{\mathcal{m}_{G}}}{\partial P_{A \alpha}} \right)_{eq} \cdot  \bra{\Omega_{Gg}} \hat{P}_{A \alpha} \ket{\Omega_{Ge}} \\
    &= \sum_k \sum_{A\alpha }  \left( \frac{\partial \bm{\mathcal{m}_{G}}}{\partial P_{A \alpha}} \right)_{eq} M_{A}  \mathcal{S}_{A \alpha,k} \ i\left(\frac{\hbar M_A \omega_k}{2} \right)^{1/2}
\end{aligned}
\end{equation}
 Plugging Eqs. \ref{eq:mu_pet} and \ref{eq:L_pet} into Eq. \ref{eq:rot_str}, we find the rotational strength for the $k^{th}$ mode to be

\begin{comment}
Using the hypervirial theorem, we rewrite $\bra{\chi_{Gg}} P_{A \alpha} \ket{\chi_{Ge}}$ for the i$^{th}$ vibrational mode as 

\begin{equation}
\begin{aligned} 
    \bra{\Psi_{Gg}}\vec{\mathcal{m}} \ket{\Psi_{Gg}}_{i} &= \sum_{A\alpha} \left( \frac{\partial}{\partial P_{A \alpha}} \bra{\Phi_{G}} \vec{\mathcal{m}} \ket{\Phi_{G}} \right)_{P=0} \frac{i M_A (E_{g} - E_{e})}{\hbar} \cdot \bra{\chi_{Gg}} (R_{A \alpha}-R^0_{A \alpha})  \ket{\chi_{Ge}}\\
    &= \sum_{A\alpha} \left( \frac{\partial}{\partial P_{A \alpha}} \bra{\Phi_{G}} \vec{\mathcal{m}}^e \ket{\Phi_{G}} + \sum_{\gamma\beta} \frac{Z_{A}e}{2M_{A}} \epsilon_{\alpha \beta \gamma}R_{A \gamma} \hat{\epsilon}_{\beta} \right) \frac{i M_A \hbar \omega_i}{\hbar} \ \mathcal{S}_{A \alpha,i} \left(\frac{\hbar}{2 \omega_i }\right)^{1/2} \\
    &= \sum_{A\alpha }\frac{i M_{A}}{2 } \frac{\partial \vec{\mathcal{m}}}{\partial P_{A \alpha}}  \mathcal{S}_{A \alpha,i} (2 \hbar \omega_i )^{1/2}
\end{aligned}
\end{equation}

\noindent Where the elements $\mathcal{S}_{A \alpha,i}$ define the i$^{th}$ normal mode according to $X_{A\alpha} = \sum_i \mathcal{S}_{A \alpha,i} Q_i$ or $\dot{X}_{A\alpha} = \sum_i \mathcal{S}_{A \alpha,i} \dot{Q}_i = \sum_i \mathcal{S}_{A \alpha,i} \frac{P_A}{M_A}?$. Note that in using the hypervirial theorem we have already assumed an infinite basis (with a fully correlated wavefuntion). We similarly expand the electric dipole moment in R, and apply the harmonic approximation to get the expression
\end{comment}

\begin{equation}
\label{eq:Rot_strv2}
\begin{aligned} 
    \mathcal{R}_{k} &=   \sum_{A \alpha, A' \alpha' } \frac{\hbar M_{A}}{2} \mathcal{S}_{A \alpha,k} \mathcal{S}_{A' \alpha',k} \left( \frac{\partial \bm{\mathcal{m}_{G}}}{\partial P_{A \alpha}} \right)_{eq}  \cdot \left( \frac{\partial \bm{\mu_{G}}}{\partial R_{A' \alpha'}} \right)_{eq} \\
\end{aligned}
\end{equation}

\subsection{\texorpdfstring{Evaluating $\bm{m}$ and $\bm{\mu}$ within Generalized Hartree-Fock Theory}{Evaluating m and mu within Generalized Hartree-Fock Theory}}\label{sec:HFexpressions}

%\subsection{Evaluating $m$ and $\mu$ within Generalized Hartree-Fock Theory}\label{sec:HFexpressions}
%Since our phase-space Hamiltonian has direct dependence on canonical momentum, calculating the electronic $\frac{\partial \bm{\hat{\mathcal{m}}}^e}{\partial P_{A\alpha}}$ is straightforward and requires only one calculation of the orbital response. Furthermore, 
Phase-space approaches break the time-reversibility of the electronic Hamiltonian and therefore necessitate complex wavefunctions whenever the nuclear momentum is nonzero. In that vein, the simplest applicable electronic structure method is generalized Hartree-Fock (GHF) theory, where the ansatz of the wavefunction is a Slater determinant where the orbitals are allowed to be complex-valued. (In the present paper, we ignore all spin-couplings, so that the orbitals do still keep $s_z$ as a good quantum number.)
Moreover, as discussed in Sec. \ref{sec:phasespaceframework} above, for nondegenerate electronic wave-functions without fine structure, $ \bP_{eq} = 0$, so that the zeroth order magnetic moment moment remains zero. With this information, we can evaluate $\bmm$ and $\bm{\mu}$ within GHF theory.

Inserting the definition for $\mmhb^e$ given in Eq. \ref{eq:medefinition} allows us to directly calculate $\frac{\partial \bm{\mathcal{m}}}{\partial \bP} $ as
(dropping the ``G'' for notational simplicity):
%Fix this so the notation is similar to before
\begin{equation}
\begin{aligned}
\label{eq:dmdPhf}
\frac{\partial \mathcal{m}_{\beta}}{\partial P_{A \alpha}}  &= \frac{\partial}{\partial P_{A \alpha}} \left(\bra{\Phi_{G}} \hat{\mathcal{m}}^e_{ \beta} \ket{\Phi_{G}} + \mathcal{m}^n_{ \beta}  \right) \\
&= \frac{-e}{2m_e} \ \frac{\partial}{\partial P_{A \alpha}} \bra{\Phi_{G}} \hat{L}^e_{\beta} \ket{\Phi_{G}} + \sum_{\gamma} \frac{Z_{A}e}{2M_{A}} \epsilon_{\alpha \beta \gamma}R_{A \gamma}   \\
&= \frac{-e }{2m_e} \sum_{\mu, \nu} \frac{\partial D_{\mu\nu}}{\partial P_{A \alpha}} \bra{\chi_{\mu}} \sum_{\gamma\delta} \epsilon_{\beta \gamma \delta} \ \hr_{\gamma} \hp_{\delta} \ket{ \chi_{\nu} } + \sum_{\gamma} \frac{Z_{A}e}{2M_{A}} \epsilon_{\alpha \beta \gamma}R_{A \gamma}  \\
%&= \frac{-e}{2m_e} \  D^{P_{A\alpha}} \cdot \left\langle  L_{e,\beta} \right\rangle
\end{aligned}
\end{equation}

\noindent %wherein we converted the calculation to a second-quantized form using the fact that $\hat{L}_{e\beta}$ is a one electron operator. 
\noindent The components of $\frac{\partial \bm{\mu}}{\partial \bR}$ are similarly given as
\begin{equation}
\begin{aligned}
\label{eq:dmudRhf}
\frac{\partial \mu_\beta}{\partial R_{A \alpha}}  &=
\frac{\partial}{\partial R_{A \alpha}} \left( \bra{\Phi_{G}}  \hat{\mu}_{\beta}^e \ket{\Phi_{G}}  + \mu^n_\beta \right) \\ &= -e \ \frac{\partial}{\partial R_{A \alpha}} \bra{\Phi_{G}} \hat{r}_{\beta} \ket{\Phi_{G}} + eZ_{A} \delta_{\beta\alpha}\\
&= -e \left( \sum_{\mu\nu} \frac{\partial D_{\mu\nu}}{\partial R_{A \alpha}} \bra{\chi_{\mu}}  \hat{r}_{\beta} \ket{ \chi_{\nu}} + D_{\mu \nu} \frac{\partial}{\partial R_{A \alpha}} \bra{\chi_{\mu}}  \hat{r}_{\beta} \ket{ \chi_{\nu}} \right) + eZ_{A} \delta_{\beta\alpha}
%&= -e \ (D^{R_{A \alpha}} \cdot \left\langle \hat{r}_\beta \right\rangle  + D \cdot \left\langle \hat{r}_\beta \right\rangle^{R_{A \alpha}} )
\end{aligned}
\end{equation}
   \\

Finally, components of $\pp{\bm{\mu}}{\bP}$ are given as

\begin{equation}
\begin{aligned}
\frac{\partial \mu_{\beta}}{\partial P_{A \alpha}}  &= \frac{\partial}{\partial P_{A \alpha}} \left(\bra{\Phi_{G}} \hat{\mu}^e_{ \beta} \ket{\Phi_{G}} + \mu^n_{\beta}  \right) \\
&= -e \sum_{\mu\nu} \frac{\partial D_{\mu\nu}}{\partial P_{A \alpha}} \bra{\chi_{\mu}}  \hat{r}_{\beta} \ket{ \chi_{\nu}}  \\
%&= \frac{-e}{2m_e} \  D^{P_{A\alpha}} \cdot \left\langle  L_{e,\beta} \right\rangle
\end{aligned}
\end{equation}

\noindent Note that at $\bP_{eq} = 0$, $\frac{\partial D_{\mu\nu}}{\partial P_{A \alpha}}$ is imaginary Hermitian, whereas the matrix $r_{\beta,\mu\nu}$ is real symmetric, and thus $\pp{\bm{\mu}}{\bP} = 0$. Expressions for calculating $\pp{D_{\mu\nu}}{R_{A\alpha}}$ and $\pp{D_{\mu\nu}}{P_{A\alpha}}$ through a CPHF phase-space approach are given in Appendix \ref{appendix:CPSCF}.

\section{Numerical Results}\label{sec:Results}

\begin{figure}[H]
    \centering
    \subfloat[\centering R-$d_2$-oxirane]{\label{fig:oxi}{\includegraphics[width=3cm]{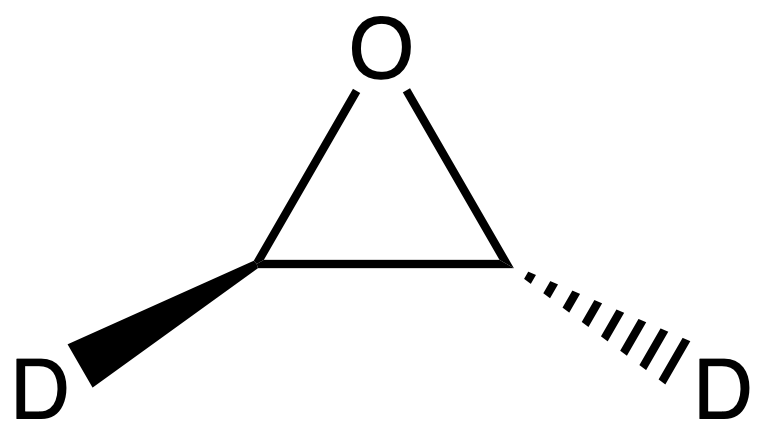}}}%
    \qquad
    \subfloat[\centering R-$d_2$-cyclopropane]{\label{fig:cycP}{\includegraphics[width=3.0cm]{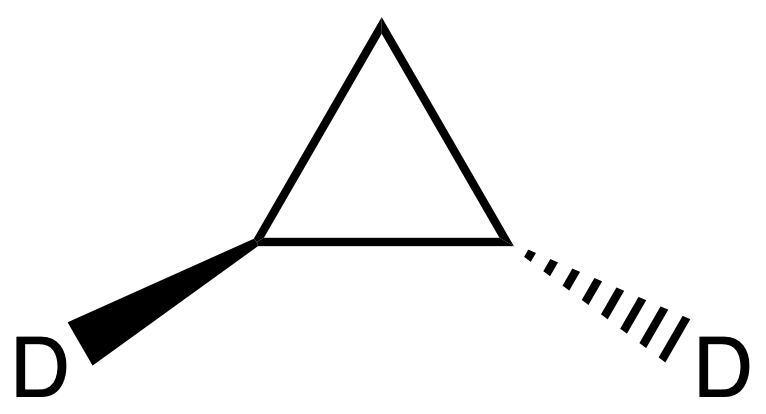}}}%
    \qquad
    \subfloat[\centering S-propylene-oxide]{\label{fig:propoxi}{\includegraphics[width=3cm]{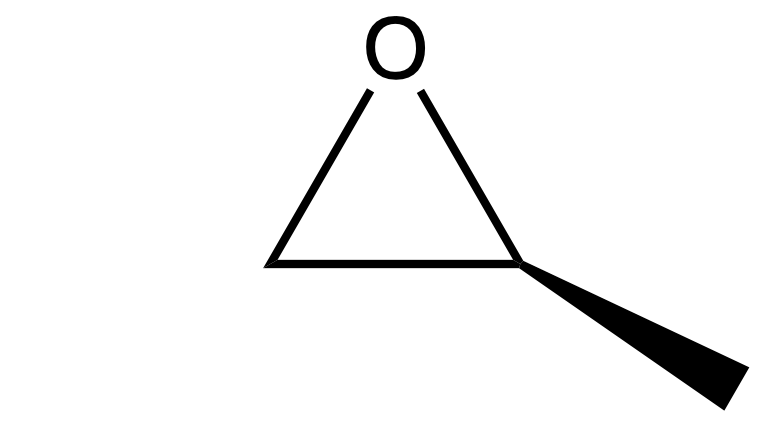} }}%
    %\subfloat[\centering R-methyl-thiirane]{\label{fig:thiirane}{\includegraphics[width=3cm]{pictures/R-methyl-thiirane.png} }}%
    \label{fig:molecules}%
\end{figure}

The theory above has been implemented within a developmental version of the Q-Chem electronic structure package.\cite{qchem4}
To benchmark the rotational strength expression from Eq. \ref{eq:Rot_strv2} above, we have modeled a set of small rigid chiral molecules: R-$d_2$-oxirane (Figure \ref{fig:oxi}), R-$d_2$-cyclopropane (Figure \ref{fig:cycP}), andS-propylene-oxide (Figure \ref{fig:propoxi}).
These molecules have been previously characterized quite sensitively by Nafie and others, \cite{expvcdcycp,expvcdoxi,expvcdmeoxi} so that we can directly compare our results vs. experimental data (with the caveat that the latter are not in the gas phase). As far as the gauge origin in concerned, all of our calculations are run with a distributed origin  gauge, whereas the MFP calculations are run with a common origin and GIAOs. For more details, see Secs. \ref{appendix:gauge}.

\subsection{VCD Spectra}

\begin{figure}[H]
\centering
\includegraphics[width=0.9\textwidth]{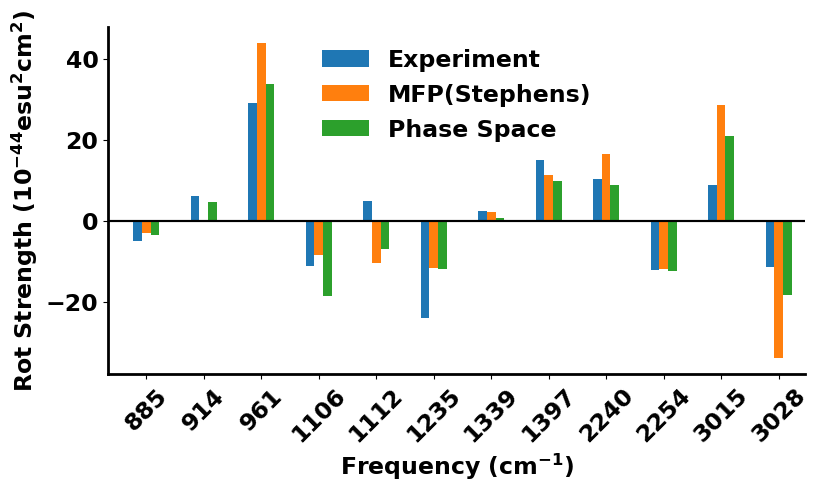}
\caption{Experimental\cite{expvcdoxi} rotational strength of R-$d_2$-oxirane vs. theoretical ones using $\Gamma$ coupling and MFP. Experimental data in $C_2Cl_4$ solution for C-H and C-D stretching modes and in $CS_2$ solution otherwise. All theoretical calculations are in vacuum (GHF/aug-cc-pvqz).  All frequencies listed on the x-axis are the experimental values. For this problem, phase-space methods would appear to outperform MFP methods. \label{fig:oxi_exp} }
\end{figure}

\begin{figure}[H]
\centering
\includegraphics[width=0.9\textwidth]{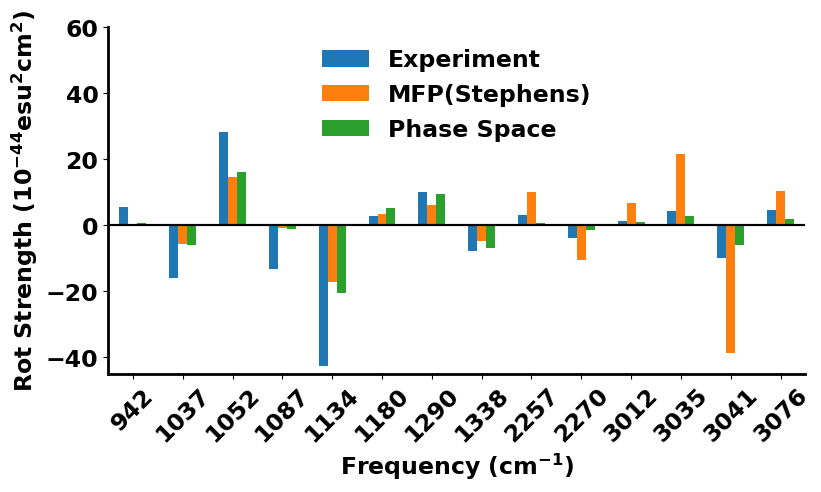}
\caption{Experimental\cite{expvcdcycp} rotational strength of R-$d_2$-cyclopropane vs. theoretical ones using $\Gamma$ coupling and MFP. Experimental data in $C_2Cl_4$ solution for C-H and C-D stretching modes and in $CS_2$ solution otherwise. All theoretical calculations are in vacuum (GHF/aug-cc-pvqz). All frequencies listed on the x-axis are the experimental values.  For this problem, phase-space methods would appear to outperform MFP methods. \label{fig:cycP_exp} }
\end{figure}

\begin{figure}[H]
\centering
\includegraphics[width=0.9\textwidth]{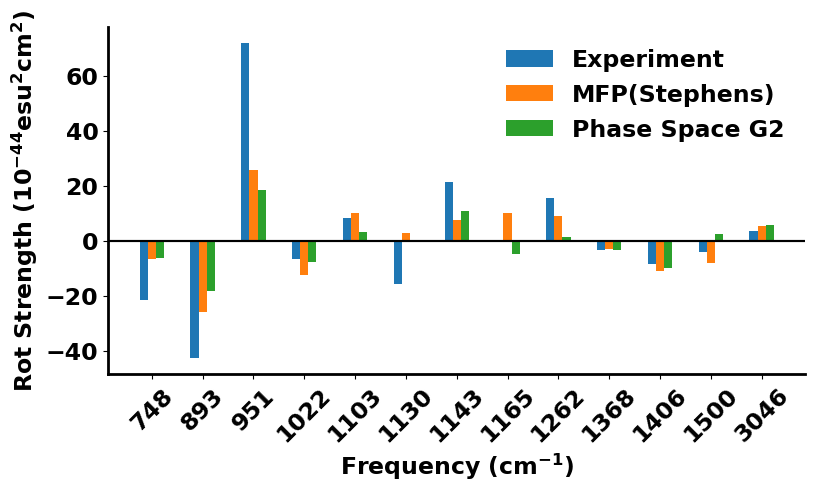}
\caption{Experimental\cite{expvcdmeoxi} rotational strength of S-propylene-oxide vs. theoretical ones using $\Gamma$ coupling and MFP. Experimental data collected in neat liquid. Mode assignments based on Ref. \citenum{assignmodemeoxi}. All theoretical calculations are in vacuum (GHF/aug-cc-pvqz). All frequencies listed on the x-axis are the experimental values. For this problem, phase-space methods would appear to underperform MFP methods. \label{fig:meoxi_exp} }
\end{figure}

\begin{figure}[H]
\centering
\includegraphics[width=0.6\textwidth]{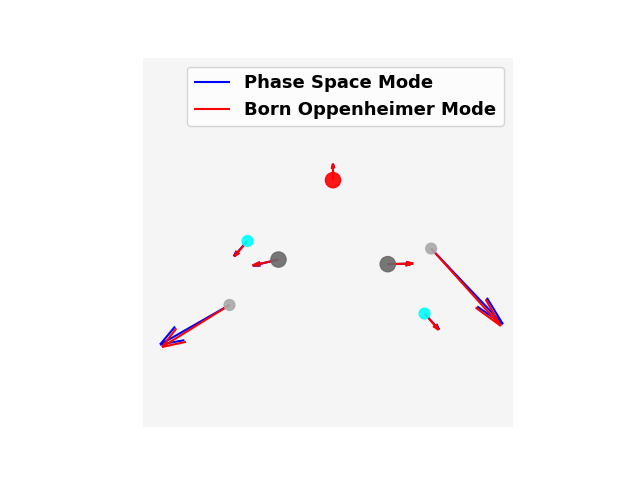}
\caption{ Oxirane 3256 $cm^{-1}$ H Sym stretch mode (3291 $cm^{-1}$ BO)  according to both a BO and phase-space Hessian (where we plot the spatial components of the phase-space mode). \label{fig:mode} }
\end{figure}

In Figs. \ref{fig:oxi_exp}-\ref{fig:meoxi_exp}, we plot $\mathcal{R}$ as calculated for the normal modes of each molecule vs. an MFP calculation and vs. experiment. 
Note that this data was gathered using the phase-space Hamiltonian in Eq. \ref{eq:finalH2} (which includes a second-derivative term). Although not shown, if one were to use the Hamiltonian in Eq. \ref{eq:eps_unstable} (without a second-derivative coupling term), the answers would be terrible -- mostly because the normal modes are nonphysical (and very different from the standard normal modes). As we highlighted above, it seems that in order to retrieve meaningful normal modes with a phase-space Hamiltonian, the kinetic energy must be positive (which is not true without the second derivative).  A visualization of a phase-space normal mode for the molecule \ref{fig:oxi} (relative to a standard normal mode) is given in Fig. \ref{fig:mode}.\\

Let us now turn to the rotatory strengths. As shown in Figs. \ref{fig:oxi_exp}-\ref{fig:meoxi_exp}, both our data and the MFP data recover the experimental data reasonably well, though we submit that phase-space methods perform better for Figures \ref{fig:oxi_exp} and \ref{fig:cycP_exp} (especially for large frequencies). 
This finding is interesting insofar as our method only includes the non-diverging portion of the electronic momentum coupling (as compared with MFP). 
Neither MFP nor our phase approach recovers the exact experimental data perfectly, for which one can propose a slew of explanations. First, on the experimental side, the data are acquired in solution, and furthermore some of the data arise between closely spaced vibrations (in the presence of mode degeneracy, ascertaining the strength of a VCD signal for a particular mode can be difficult).
Second, on the theoretical side, we note that we have used a GHF calculation and entirely ignored direct electron-electron correlation. We have also sampled only a single nuclear geometry per molecule, and we have ignored 
all anharmonic effects \cite{VCDsolvent,VCDHbond,fuse2022anharmonic}. 
Notwithstanding all of these limitations, the rough accuracy of the data presented in Fig. \ref{fig:oxi_exp}-\ref{fig:meoxi_exp} highlights the fact that even a crude phase-space approach can recover some very sensitive optical signals. For the raw data behind these graphs,  see that tables in Appendix \ref{appendix:tables}. 

%It should be noted that errors originating from deviations from normal modes are a flaw of the harmonic analysis at our level of theory rather than a flaw in our methodology. Our ability to achieve agreement, even in sign, lends substantial credibility to our approach.\\

Before concluding this paper, a few words are appropriate regarding the non-BO part of the W term in Eq. \ref{eq:tot_nuc_hess}. Note that the concept of effective masses is very old and has been standardized in just about all spectroscopy books. Here, however, we have found that with a phase-space electronic Hamiltonian, we must replace the diagonal inverse mass tensor with a non-diagonal ``inverse mass'' tensor (which must be diagonalized in order to recover spectra, see Sec \ref{sec:PSModes}).
One can ask: For such a non-diagonal tensor, what is the meaning of the on-diagonal elements? How close do the values resemble the inverse masses of the nuclei?

To that end, in Tables \ref{tbl:effmass}-\ref{tbl:Wdiag}, we list both  the inverse of the eigenvalues of $W$ and the inverse of the diagonal elements of $W$. In some sense, these values must correspond to ``effective masses'' insofar as they have units of mass, but these masses depend on direction and the eigenvectors of $W$ are delocalized. While we cannot easily make any clear conclusions regarding the values of these masses, there is one obvious conclusion from this data: namely, all of the masses are {\em larger} than the raw masses (usually by $10^{-2}$ amu). In this regard, experimentalists have pointed out that for diatomics,\cite{kutzelnigg:2007:mp, gross:2017:prx:born_oppenheimer_mass} according to high-resolution spectroscopy, when constructing $\hat{H}_{vib}$, the masses of nuclei should be set to the mass of the atom (which includes nucleus plus electrons). For a carbon atom with 6 electrons, one would therefore expect an increased carbon mass of $6/1 822.9 \approx .003$ amu. For the molecules considered hitherto, we predict even larger masses (around tenfold, see Table \ref{tbl:effmass}) -- though a clear interpretation of these masses is difficult since we also have off-diagonal components of $W$ and these molecules are not as small as diatomics.

\begin{table}[H]
\caption{The inverses of the eigenvalues of $W$ in Eq. \ref{eq:tot_nuc_hess}  in a.m.u. (which are one estimate of a dressed ``effective mass'').}\label{tbl:effmass} 
\begin{tabular}{ccc}
\hline \hline
\multicolumn{1}{l}{\textbf{R-d2-oxirane}}  	&	\multicolumn{1}{l}{\textbf{R-d2-cyclopropane}}	&	\multicolumn{1}{l}{\textbf{S-propylene-oxide}}	\\ \hline
1.0309	&	2.0338	&	16.041	\\
1.0307	&	2.0335	&	16.0249	\\
1.0164	&	2.0224	&	16.0151	\\
1.0166	&	2.022	&	12.0318	\\
1.0195	&	2.0246	&	12.0534	\\
1.0199	&	2.0246	&	12.0344	\\
2.0369	&	1.0281	&	12.0366	\\
2.0367	&	1.028	&	12.0503	\\
2.0226	&	1.0274	&	12.0488	\\
2.0228	&	1.027	&	12.0425	\\
2.0257	&	1.013	&	12.0442	\\
2.0261	&	1.014	&	12.0434	\\
16.0382	&	1.0169	&	1.0109	\\
16.0237	&	1.0193	&	1.0302	\\
16.013	&	1.0189	&	1.0297	\\
12.0599	&	1.0181	&	1.0292	\\
12.0415	&	1.0186	&	1.0285	\\
12.0528	&	1.0184	&	1.0132	\\
12.045	&	12.0266	&	1.0137	\\
12.048	&	12.0265	&	1.0264	\\
12.0466	&	12.0315	&	1.026	\\
	&	12.0355	&	1.0155	\\
	&	12.0462	&	1.0166	\\
	&	12.042	&	1.0216	\\
	&	12.042	&	1.021	\\
	&	12.0444	&	1.0207	\\
	&	12.0444	&	1.0203	\\
	&		&	1.0191	\\
	&		&	1.0193	\\
	&		&	1.0193	\\
\hline
\end{tabular}
\end{table}

\begin{table}[H]
\caption{Inverse diagonal elements of $W$ in Eq. \ref{eq:tot_nuc_hess} (which are another estimate of a dressed ``effective mass''). }\label{tbl:Wdiag} 
\begin{tabular}{cccccc}
\hline \hline
\multicolumn{2}{l}{\textbf{R-d2-oxirane}}  			&  \multicolumn{2}{l}{\textbf{R-d2-cyclopropane}}				&  \multicolumn{2}{l}{\textbf{S-propylene-oxide}}			\\ \hline	
C	&	12.0470	&	H	&	1.0214	&	C	&	12.0411	\\
C	&	12.0492	&	H	&	1.0171	&	C	&	12.0369	\\
C	&	12.0497	&	H	&	1.0233	&	C	&	12.0409	\\
C	&	12.0470	&	C	&	12.0369	&	C	&	12.0470	\\
C	&	12.0492	&	C	&	12.0372	&	C	&	12.0483	\\
C	&	12.0497	&	C	&	12.0382	&	C	&	12.0460	\\
H	&	1.0217	&	H	&	1.0182	&	H	&	1.0249	\\
H	&	1.0263	&	H	&	1.0203	&	H	&	1.0189	\\
H	&	1.0189	&	H	&	1.0233	&	H	&	1.0215	\\
H	&	1.0217	&	H	&	1.0182	&	H	&	1.0189	\\
H	&	1.0263	&	H	&	1.0203	&	H	&	1.0229	\\
H	&	1.0189	&	H	&	1.0233	&	H	&	1.0252	\\
O	&	16.0235	&	C	&	12.0374	&	H	&	1.0202	\\
O	&	16.0129	&	C	&	12.0367	&	H	&	1.0194	\\
O	&	16.0380	&	C	&	12.0382	&	H	&	1.0287	\\
D	&	2.0279	&	H	&	1.0182	&	C	&	12.0420	\\
D	&	2.0324	&	H	&	1.0203	&	C	&	12.0411	\\
D	&	2.0251	&	H	&	1.0233	&	C	&	12.0396	\\
D	&	2.0279	&	C	&	12.0369	&	H	&	1.0194	\\
D	&	2.0324	&	C	&	12.0372	&	H	&	1.0237	\\
D	&	2.0251	&	C	&	12.0382	&	H	&	1.0160	\\
	&		&	D	&	2.0244	&	H	&	1.0196	\\
	&		&	D	&	2.0265	&	H	&	1.0214	\\
	&		&	D	&	2.0295	&	H	&	1.0182	\\
	&		&	D	&	2.0276	&	H	&	1.0184	\\
	&		&	D	&	2.0233	&	H	&	1.0171	\\
	&		&	D	&	2.0295	&	H	&	1.0261	\\
	&		&		&		&	O	&	16.0250	\\
	&		&		&		&	O	&	16.0358	\\
	&		&		&		&	O	&	16.0198	\\
\hline
\end{tabular}
\end{table} 

\section{Discussion and Conclusions}\label{sec:Conclusion}

\noindent We have presented a phase-space electronic structure approach for calculating VCD signals within GHF theory and compared our results against experiment for three model test cases. Our results demonstrate fairly conclusively that such a phase-space approach -- which entirely avoids the BO approximation -- is valid and can yield insight into nuclear-electronic coupling that is not available within a BO approximation.  Our implementation in a developmental version of QChem \cite{qchem4} is currently quite slow, but in principle there is no reason such a phase-space approach should be much slower than standard electronic structure approaches, with the main caveat being the need for complex-valued (rather than real-valued) multiplication.

Interestingly, one finds that our results in Figs. \ref{fig:oxi_exp}-\ref{fig:meoxi_exp} are comparable with the results of the standard VCD technique, magnetic field perturbation theory. In fact, for Figs. \ref{fig:oxi_exp} and \ref{fig:cycP_exp}, one can easily argue that phase-space approaches outperform MFP theory; though the opposite can be argued for Fig. \ref{fig:meoxi_exp}. While MFP theory is reviewed below (in Appendix \ref{appendix:MFP}), it is worth mentioning that MFP is formally a higher-order level calculation than phase-space approaches because the latter accounts only for how electronic orbitals are dragged by nuclei (i.e. the ETF component of the derivative coupling in Eq. \ref{eq:decompose_d}), whereas the former includes the entire orbital response $\left(\left|\frac{\partial \Phi_I}{\partial R_{A\alpha}}\right> \text{, i.e. the entire derivative coupling}\right)$. Thus, one would imagine that MFP theory would match better with Shenvi's phase-space formalism (in Eq. \ref{eq:shenvi}) than would ours.

Thus one must wonder: on the one hand, does the simple phase-space electronic structure Hamiltonian in Eq. \ref{eq:finalH1} perform so well in Figs. \ref{fig:oxi_exp}-\ref{fig:meoxi_exp} because the relevant molecules are adiabatically very stable and far from any avoided crossing to another electronic state, but we would not recover such strong results for a more mixed-valence compound? Or on the other hand, because we argued above that Shenvi's phase-space approach would face limitations near a conical intersection, is it possible that MFP theory will become less accurate near a crossing while our phase-space approach will remain stable? Or perhaps, of course, neither approach will be stable. Ultimately, this thought experiment is a strong reason to run future VCD calculations for more interesting molecules where there is already some ground-state mixing present at equilibrium. \\

Looking forward, in the future we should also confront another key question that inevitably arises in any VCD calculation, namely the dependence  on the chosen origin for the perturbative magnetic field. More details about the choice of origin are given in Appendix \ref{appendix:gauge}, but the take home-points are as follows:
\begin{itemize}
    \item On the one hand,  the choice of origin should not affect MFP theory in the limit of an infinite (complete) basis. For an incomplete basis, one usually runs MFP calculations with GIAOs (which automatically eliminate origin dependence). Indeed, the MFP data in Figs. \ref{fig:oxi_exp}-\ref{fig:meoxi_exp} were generated with GIAOs.
    \item On the other hand, phase-space calculations using the formalism in Sec. \ref{sec:rotstr} are  sensitive to the choice of gauge origin (no matter how big the basis). For this reason, in order to mitigate the dependence on the gauge origin, we ran the phase-space calculations with a distributed origin gauge.  That being said, we still found that we required a large basis for the calculations to match experiment.
\end{itemize}
    With this background in mind, if we seek to improve our phase space approach, one means to address this dependence on origin is to implement the formalism above and evaluate the electronic energy using GIAOs (that eliminate any gauge dependence for $E_{PS}$) and to perform the substitution:
\begin{equation}
\begin{aligned}
\frac{\partial {\mathcal{m}}^e_{ \beta}}{\partial P_{A \alpha}}  &=  -\frac{\partial}{\partial P_{A \alpha}} \left(\frac{\partial E_{PS}}{\partial B_\beta} \right) = -\frac{\partial}{\partial B_\beta} \left( \frac{\partial E_{PS}}{\partial P_{A \alpha}}\right) 
\end{aligned}
\end{equation}
Because $E_{PS}$ is invariant to gauge origin by construction,
the resulting VCD signal must then be independent of origin as well. Ideally, such an approach should also converge easily for smaller basis sets, though this remains to be proven. Clearly, implementing such a formalism is one of our most immediate goals in the near term.

Finally, let us turn to the implications of the present results insofar as the future of phase-space electronic structure Hamiltonians. We can imagine two important avenues. First, the presence of the extra terms in a phase-space electronic Hamiltonian (Eq. \ref{eq:finalH1}) as compared with a BO electronic Hamiltonian (Eq. \ref{eq:standard}) should have consequences for dynamics and not just for spectroscopy. Most importantly, we imagine that consequences may well arise in the context of electron transfer with spin degrees of freedom. As discussed in Sec. \ref{sec:Outline}, normal BO dynamics do not conserve the total angular or linear momentum of a composite system of nuclei and electrons (and potentially spins). In the context of electron transfer, however, one can easily imagine a paradox: When an electron moves from donor to acceptor, there must be some change in electronic momentum, which must be balanced by another form of momentum (for total momentum conservation). In the context of linear momentum, that extra electronic linear momentum must be balanced by nuclear linear momentum; in the context of angular momentum, however, that extra electronic angular momentum can be balanced by either spin or nuclear angular momentum. Thus, there is the possibility that the current phase-space electronic structure picture will lead to spin-dependent electron transfer dynamics and function as one potential framework for understanding recent experiments in CISS.\cite{Naaman2012,Naaman2019,Evers2022,Das2022,Fransson2023} Note that, in Secs. \ref{sec:PSModes} and \ref{sec:rotstr} above (and Appendix \ref{appendix:MFP} below), we have consistently used the fact that, at equilibrium, $\bP_{eq} = \dot{\bR}_{eq} = 0$, but this is guaranteed to be true only in the absence of SOC; thus, in the presence of SOC, one can indeed anticipate new and exciting physics.\\

Second, because the phase-space electronic structure Hamiltonian depends on nuclear momentum (or really velocity), one must wonder if the current formalism will offer a new approach towards modeling superconductivity. After all, it has been well documented that for 
 many superconductors,\cite{isotopeSC1,isotopeSC2} the critical temperature depends sensitively on phonons and, in particular, the mass of the nuclear isotope. Furthermore, intuitively, the extra term in the kinetic energy in Eq. \ref{eq:maybe} reflects the energetic consequences of the fact that whenever a nuclear charge moves, that charge carries an orbital. Thus, whenever a nucleus is attracted to and moves towards an electron, by default, that motion induces a slight attraction between the electron carried by the nucleus and the electron to which the nucleus is attracted. This state of affairs would seem to capture some of the essence of superconductivity, and further work in this vein would appear appropriate. One must wonder if, e.g., the presence of extra terms in a phase-space Hamiltonian can strongly change the nature of the electron-electron correlation problem and, e.g., induce Cooper pairing\cite{Gao2023}.

Overall, phase-space electronic Hamiltonians would appear to be a very promising technique for future explorations in quantum chemistry, quantum dynamics, and beyond.

\section{Acknowledgments}
This work was supported by the U.S. Air Force Office of Scientific Research (USAFOSR)
under Grant Nos. FA9550-23-1-0368 and FA9550-18-1-420. YS acknowledges support from NSF grant CHE-2102071.

\appendix

\section{Dependence on the Gauge Origin} \label{appendix:gauge}

Quite generally, CD spectra (and we find especially VCD spectra) can depend sensitively on where one chooses the origin of the magnetic field operator in Eq. \ref{eq:medefinition}, \ref{eq:mndefinition}. 
In the interest of transparency, it is worth noting that our naive phase-space method will suffer from a magnetic origin problem as well; this stands in contrast to MFP, SOS, and NVP calculations, where one can show that given an infinite basis, the rotational strength will be gauge invariant. To explain why this is so, we will now review Stephen's original work on the gauge dependence of the rotational strength \citenum{stephens1987gauge} and comment on its implications for phase-space methods.

To begin our discussion, consider how $\bm{\mathcal{m}}$ changes under the origin transformation $\bm{\mathcal{O'}} = \bm{\mathcal{O}} + \bm{Y}$. Unsurprisingly, $\bm{\mathcal{m}}$ transforms much like angular momentum. 
\begin{equation}
\begin{aligned}
(\bm{\mathcal{m}}^e)^{\mathcal{O}} = (\bm{\mathcal{m}}^e)^\mathcal{O'} -\frac{e}{2m_e} (\bm{Y} \times \expval{\bm{\hat{p}}})\\
(\bm{{\mathcal{m}}}^n)^{\mathcal{O}} = (\bm{\mathcal{m}}^n)^{\mathcal{O'}} + \sum_B \frac{Z_{B}e}{2M_{B}} (\bm{Y} \times \bm{P})
\end{aligned}
\end{equation}

\noindent Calculating $\frac{\partial \mathcal{m}_\beta}{\partial P_{A\alpha}}$ as before, we find

\begin{eqnarray}
\label{eq:Orgn_chg}
\frac{\partial \mathcal{m}_\beta^{\mathcal{O}}}{\partial P_{A\alpha}} &=&\frac{\partial \mathcal{m}_\beta^{\mathcal{O}'}}{\partial P_{A\alpha}} + \frac{-e}{2m_e} \sum_{\gamma \delta} \epsilon_{\beta \gamma \delta} Y_{\gamma} \frac{\partial \expval{\hat{p}_\delta}}{\partial P_{A \alpha}} + \sum_{B \gamma \delta} \frac{Z_{B}e}{2M_{B}}  \epsilon_{\beta \gamma \delta} Y_{\gamma} \frac{\partial  P_{B \delta} }{\partial P_{A \alpha}} \\
&=&\frac{\partial \mathcal{m}_\beta^{\mathcal{O}'}}{\partial P_{A\alpha}} + \frac{1}{2 M_A } \sum_{\gamma \delta}\left( - e\epsilon_{\beta \gamma \delta} Y_{\gamma}  \frac{\partial \expval{\hr_\delta}}{\partial R_{A \alpha}} + eZ_{A} \epsilon_{\beta \gamma \delta} Y_{\gamma} \delta_{\delta \alpha} \right) \\
&= & \frac{\partial \mathcal{m}_\beta^{\mathcal{O}'}}{\partial P_{A\alpha}}  + \frac{1}{2 M_A } \sum_{\gamma \delta}\epsilon_{\beta \gamma \delta} Y_{\gamma}  \frac{\partial\mu_\delta}{\partial R_{A \alpha}}
\end{eqnarray}

\noindent In the first step, we have used the equivalent of Ehrenfest's theorem highlighted by  Nafie in the context of BO theory, namely the fact that $\expval{\bm{\hat{p}}} = m (d\expval{\bm{\hat{r}}}/dt)$ , which allows us to write

\begin{equation}
\begin{aligned}
\label{eq:Nafiethrm}
\frac{\partial \expval{\bm{\hp}}}{\partial P_{A\alpha}} &= m_e \frac{\partial}{\partial P_{A\alpha}} \frac{\partial \expval{\bm{\hr}}}{\partial t}\\
&= \frac{\partial}{\partial{P_{A\alpha}}} \sum_{A\alpha} \frac{m_e}{M_A} \frac{\partial\expval{\bm{\hr}}}{\partial R_{A\alpha}} M_A \dot{R}_{A\alpha} \\
&=\frac{m_e}{M_A} \frac{\partial\expval{\bm{\hr}}}{\partial R_{A\alpha}}
\end{aligned}
\end{equation}
(see Sec. 3 of Ref. \citenum{coraline24gamma}). 

Note, however, that in the context of a phase-space electronic Hamiltonian with a $\Gamma$ coupling, this condition is only approximately satisfied.\cite{coraline24gamma} That being said, if we  invoke Nafie's approximation and we insert Eq. \ref{eq:Orgn_chg} into Eq. \ref{eq:Rot_strv2} for the rotational strength, we find:

\begin{equation}
\begin{aligned}
\label{eq:gauge}
    \mathcal{R}_{k}^{\mathcal{O}'} &=   \sum_{A \alpha, A' \alpha' } \sum_\beta \frac{\hbar M_{A}}{2} \mathcal{S}_{A \alpha,k} \mathcal{S}_{A' \alpha',k}  \frac{\partial \mu_\beta}{\partial R_{A' \alpha'}} 
    \frac{\partial \mathcal{m}_\beta^{\mathcal{O}'}}{\partial P_{A \alpha}}  \\
    &= \sum_{A \alpha, A' \alpha' } \sum_\beta \frac{\hbar M_{A}}{2} \mathcal{S}_{A \alpha,k} \mathcal{S}_{A' \alpha',k}  \frac{\partial \mu_\beta}{\partial R_{A' \alpha'}} \left( \frac{\partial \mathcal{m}_\beta^{\mathcal{O}}}{\partial P_{A \alpha}} + \frac{1}{2 M_A } \sum_{\gamma \delta}\epsilon_{\beta \gamma \delta} Y_{\gamma}  \frac{\partial\mu_\delta}{\partial R_{A \alpha}}\right)\\
    &= \mathcal{R}_{k}^{\mathcal{O}} + \sum_{A \alpha, A' \alpha' } \sum_\beta  \hbar \mathcal{S}_{A \alpha,k} \mathcal{S}_{A' \alpha',k} \sum_{\gamma \delta}\epsilon_{\beta \gamma \delta} Y_{\gamma}  \frac{\partial \mu_\beta}{\partial R_{A' \alpha'}} \frac{\partial\mu_\delta}{\partial R_{A \alpha}} \\
\end{aligned}
\end{equation}

\noindent Note that when $\bm{Y}$ is some constant vector, $\mathcal{R}_{k}^{\mathcal{O}'} = \mathcal{R}_{k}^{\mathcal{O}}$.

\begin{equation}
\begin{aligned}
    \mathcal{R}_{k}^{\mathcal{O}'} &=  \mathcal{R}_{k}^{\mathcal{O}} + \sum_{\gamma \delta\beta} \epsilon_{\beta \gamma \delta} Y_{\gamma} \frac{\partial \mu_\beta}{\partial Q_{j}}\ \frac{\partial \mu_\delta}{\partial Q_{j}} \\
    &= \mathcal{R}_{k}^{\mathcal{O}} + \left(\bm{Y} \cdot \frac{\partial \bm{\mu}}{\partial Q_{j}} \times \frac{\partial \bm{\mu}}{\partial Q_{j}} \right) = \mathcal{R}_{k}^{\mathcal{O}} 
\end{aligned}
\end{equation}

\begin{comment}

\begin{equation}
\begin{aligned}
    &= \mathcal{R}_{k}^{\mathcal{O}} +  \hbar \sum_{\gamma \delta\beta}\epsilon_{\beta \gamma \delta} Y_{\gamma}  \frac{\partial \mu_\beta}{\partial Q_{j}}\frac{\partial\mu_\delta}{\partial Q_k}\\
 &=\sum_{\gamma \delta\beta} \epsilon_{\beta \gamma \delta} Y_{\gamma} \frac{\partial \mu_\beta}{\partial Q_{j}}\ \frac{\partial \mu_\delta}{\partial Q_{j}} = \left(\bm{Y} \times \frac{\partial \bm{\mu}}{\partial Q_{j}} \times \frac{\partial \bm{\mu}}{\partial Q_{j}} \right) = 0\\
\end{aligned}
\end{equation}

\end{comment}

Now, the theory above is exact provided that Nafie's relationship in Eq. \ref{eq:Nafiethrm} holds. More generally, however, such a relationship holds only in an infinite (complete) basis. Therefore, to the extent that one always seeks to reduce dependence on basis, in the literature one finds the notion of using a distributed gauge origin (DO), rather than a common, fixed origin (CO). Here, one imagines looping over all atoms and shifting the origin to match up with the atom under consideration.  Thus, in Eq. \ref{eq:Orgn_chg} above, if one is considering atom $A$, one chooses $\bY = \bR_A$.  This then implies that:  

\begin{equation}
\begin{aligned}
\left( \frac{\partial \mathcal{m}_\beta}{\partial P_{A\alpha}}\right)^{CO} &=&\left(\frac{\partial \mathcal{m}_\beta}{\partial P_{A\alpha}}\right)^{DO} + \frac{-e}{2m_e} \sum_{\gamma \delta} \epsilon_{\beta \gamma \delta} R_{A\gamma} \frac{\partial \expval{\hat{p}_\delta}}{\partial P_{A \alpha}} + \sum_{B \gamma \delta} \frac{Z_{B}e}{2M_{B}}  \epsilon_{\beta \gamma \delta} R_{A\gamma} \frac{\partial  P_{B \delta} }{\partial P_{A \alpha}}
\end{aligned}
\end{equation}

Thereafter, if one again invokes Nafie's theorem in Eq. \ref{eq:Nafiethrm}, one can immediately simply the second term and repeat the procedure above to find

\begin{eqnarray}
\label{eq:RDO}
\mathcal{R}_{k}^{CO} & =& \mathcal{R}_{k}^{DO} + \sum_{A \alpha, A' \alpha' } \sum_\beta  \hbar\mathcal{S}_{A \alpha,k} \mathcal{S}_{A' \alpha',k} \sum_{\gamma \delta}\epsilon_{\beta \gamma \delta} R_{A\gamma}  \frac{\partial \mu_\beta}{\partial R_{A' \alpha'}} \frac{\partial\mu_\delta}{\partial R_{A \alpha}} \label{eq:do2}
%&= & \sum_{A \alpha, A' \alpha' } \sum_\beta \frac{\hbar M_{A}}{2} \mathcal{S}_{A \alpha,k} \mathcal{S}_{A' \alpha',k}  \frac{\partial \mu_\beta}{\partial R_{A' \alpha'}} \left(\frac{\partial \mathcal{m}_\beta}{\partial P_{A \alpha}} \right)^{CO} \\
\end{eqnarray}
where
\begin{eqnarray}
\label{eq:RDO2}
\mathcal{R}_{k}^{DO} &= & \sum_{A \alpha, A' \alpha' } \sum_\beta \frac{\hbar M_{A}}{2} \mathcal{S}_{A \alpha,k} \mathcal{S}_{A' \alpha',k}  \frac{\partial \mu_\beta}{\partial R_{A' \alpha'}} 
    \left(\frac{\partial \mathcal{m}_\beta}{\partial P_{A \alpha}} \right)^{DO} 
\end{eqnarray}
The second term in Eq. \ref{eq:do2} (coined the $L\cdot P$ term by Stephens\cite{stephens1987gauge}) is well known. The explicit gauge origin dependence is hidden in $\mathcal{R}_{k}^{DO}$ and, in practice, evaluating this expression often gives much better results for an incomplete basis than simply picking the gauge origin as the center of nuclear charge. Indeed, we  have found that without this expression, our results can be very sensitive to the origin. Thus, all data gathered in this paper (for Figs. \ref{fig:oxi_exp}-\ref{fig:meoxi_exp}) used Eq. \ref{eq:RDO}.

\section{Nuclear Momentum CPSCF}
\label{appendix:CPSCF}

In this section we will present the equations needed to solve for the orbital response derivatives for the Hamiltonian given in Eq. \ref{eq:finalH2} (i.e. how to calculate $\pp{\bmu}{\bR}$, $\pp{\bmm}{\bP}$  in Eqs. \ref{eq:dmudRhf}, \ref{eq:dmdPhf}, and the elements of $\bW$, $\bK$, $\bY$ in Eqs. \ref{eq:dEdPdP}, \ref{eq:dEdPdR}, \ref{eq:dEdRdR}). Before embarking on such a derivation, we note that we use the following notation below for molecular orbitals:
\begin{itemize}
    \item  Roman letters a,b denote virtual molecular orbitals
    \item Roman letters i,j denote occupied molecular orbitals
    \item Roman letters p,q,n denote all molecular orbitals 
\end{itemize}

To begin our derivation, we suppose that there exists a unitary transformation of the orbital coefficients which transforms the orbital coefficients from their current value to a different value as determined some small perturbation in the variable of interest X. This unitary transformation can be written as

\begin{equation}
\label{eq:U}
c_{\mu q}^{[X]}=\sum_{n}^{n_{orb}}c_{\mu  n}U_{nq}^{[X]} 
\end{equation}

\begin{comment}

\begin{equation}
\frac{\partial \chi_J}{\partial \textbf{[X]}} = \sum_\mu \left( \frac{\partial \chi_\mu}{\partial \textbf{X}}c_{\mu j}  + \frac{\partial c_{\mu j}}{\partial \textbf{X}} \chi_\mu \right)
\end{equation}

\noindent Within the Hartree Fock framework, the $\frac{\partial c_{\mu j}}{\partial \textbf{X}}$ can be expressed in terms of the coefficients of all molecular orbitals, both occupied and unoccupied

\begin{equation}
\frac{\partial c_{\mu j}}{\partial \textbf{[X]}} = \sum_i U_{ij}^[X] c_{\mu i}^0
\end{equation}

\end{comment}

\noindent As shown  by Ref. \citenum{MODeriv}, $U^{[X]}$ can be determined by differentiating the Fock equation w.r.t X and retaining the first order terms. Rather than solve for the entire $U^{[X]}$, because only the occupied virtual block is ultimately relevant\cite{MODeriv}, we opt to use the antisymmetrized occupied virtual $ \tilde{U}^{[X]} = \frac{1}{2}(U^{[X]} - U^{[X]\dagger})$. Carrying out the CPSCF procedure yields the matrix equation

\begin{equation}
A \tilde{U}^{[X]} = B^{[X]}
\end{equation}

\noindent where $A$ is independent of the perturbation and is given as 
\begin{equation}
(A\tilde{U}^{[X]})_{ai} = \sum_{b j}\left[F_{a b} \delta_{i j}-F_{j i} \delta_{a b}+(a i \| j b)\right] \frac{1}{2} (U_{bj}^{[X]}-U_{jb}^{[X]*}) +\sum_{b j}(a i \| b j) \frac{1}{2} (U_{bj}^{[X]*}-U_{jb}^{[X]})
\end{equation}

\noindent The form of $B^{[X]}$ does depend on the perturbation, but can generally be written as
\begin{equation}
B^X_{ai} = -\bar{F}_{ai}^{[X]} + \frac{1}{2} \sum_{j} \bar{S}_{aj}^{[X]} F_{ji} + \frac{1}{2} \sum_{b} F_{ab} \bar{S}_{bi}^{[X]} + \frac{1}{2} \sum_{jn} \Big[ (ai \| jn) \bar{S}_{nj}^{[X]} + (ai \| nj) \bar{S}_{jn}^{[X]} \Big]
\end{equation}
\noindent where bar matrices indicate the derivative contains only the explicit and atomic orbital contribution(i.e. no MO coefficient derivatives). In the case where $\frac{\partial \chi}{\partial X} = 0$ (which is true for $\textbf{X} = P_{A\alpha}$), this result simplifies to 
%\begin{equation}
%\bar{S}_{p q}^{[X]}=\sum_{\mu v \tau \kappa} c_{\mu p}^{\tau *} S_{\mu v}^{[X]} c_{v q}^{\kappa} \delta_{\tau \kappa}
%\end{equation}

\begin{equation}
B_{ai}^{[P_{A\alpha}]} = -\bar{F}_{ai}^{[P_{A\alpha}]} = -\sum_{\mu\nu\tau\kappa}c^{\tau*}_{\mu a}\bra{\chi_\mu} \frac{\partial h}{\partial P_{A\alpha}}\ket{\chi_\nu} c^{\kappa}_{\nu i} = \sum_{\mu\nu\tau\kappa}c^{\tau*}_{\mu a} \left( i\hbar \frac{\Gamma^{A\alpha}_{\mu\nu}}{M_{A}} \right) c^{\kappa}_{\nu i}
\end{equation}

\noindent Solving these equations corresponds to finding the solutions to the matrix equation

\begin{equation}
\label{eq:CPSCF}
    \begin{bmatrix}
F_{a b} \delta_{i j}-F_{j i} \delta_{a b}+(a i \| j b) &  (a i \| b j)\\
(a i \| b j)^{*} & F^{*}_{a b} \delta_{i j}-F^{*}_{j i} \delta_{a b}+(a i \| j b)^{*}
\end{bmatrix}
\begin{bmatrix}
\frac{1}{2} (U_{bj}^{[X]}-U_{jb}^{[X]*}) \\
\frac{1}{2} (U_{bj}^{[X]*}-U_{jb}^{[X]})
\end{bmatrix}
=
\begin{bmatrix}
B^{[X]}_{ai}\\
B^{[X]*}_{ai}
\end{bmatrix}
\end{equation}

\noindent Finally, using the identity that $U_{ja}^{[X]*}+ U_{aj}^{[X]}+ \bar{S}_{aj}^{[X]}=0$, density derivatives are calculated as 

\begin{equation}
    D^{\tau\kappa,[X]}_{\mu\nu} 
     = \frac{1}{2}\left[\sum_{ia} \left(c^{\tau}_{\mu a}(U^{[X]}_{ai} - U^{[X]*}_{ia}) c^{\kappa*}_{\nu i}  + c^{\tau}_{\mu i}(U^{[X]*}_{ai} -U^{[X]}_{ia}) c^{\kappa*}_{\nu a}\right)  - \sum_{n\sigma\lambda\eta} \left(c^{\tau}_{\mu n}c^{\eta*}_{\sigma n}S^{[X]}_{\sigma\lambda}D^{\eta\kappa}_{\lambda \nu}  - D^{\tau\eta}_{\mu \sigma} S^{[X]}_{\sigma\lambda}c^{\eta}_{\lambda n} c^{\kappa*}_{\nu n}\right)\right]
\end{equation}

\section{Computing K, W, and Y in Eq. \ref{eq:tot_nuc_hess} : Expanding the GHF Phase Space Hamiltonian Energy to Second Order}\label{appendix:DiagonalHess}
In this section of the Appendix, we will evaluate the matrices $\bK, \bW$, and $\bY$ in Eq. \ref{eq:tot_nuc_hess}.
Quite generally, the energy of a GHF wavefunction is given by the Eq. \ref{eq:eps_stable}. Carrying out the Hartree Fock derivatives in the standard fashion (and noting that overlap momentum derivatives vanish $S^{[P]} =0$), yields the relatively simple expressions:

\begin{eqnarray}
\label{eq:dEdPdR}
\frac{\partial^2 E_{PS}}{\partial P_{A\alpha} \partial R_{B\beta}} &=& -i\hbar \sum_{\mu \nu \tau} \frac{\Gamma_{\mu \nu}^{A\alpha}}{M_A}(D^{\tau \tau}_{\mu \nu})^{[R_{B\beta}]} - i\hbar \frac{(\Gamma_{\mu \nu}^{A\alpha})^{[R_{B\beta}]}}{M_A}D_{\mu \nu}^{\tau \tau}\\
\label{eq:dEdPdP}
\frac{\partial^2 E_{PS}}{\partial P_{A\alpha} \partial P_{B\beta}} &=& \frac{1}{M_A}\delta_{AB}\delta_{\alpha\beta} -i\hbar \sum_{\mu \nu \tau} \frac{\Gamma_{\mu \nu}^{A\alpha}}{M_A}(D^{\tau \tau}_{\mu \nu })^{[P_{B\beta}]} \\
\label{eq:dEdRdR}
\frac{\partial^2 E_{PS}} {\partial R_{A\alpha} \partial R_{B\beta}} &=&   \sum_{\mu \nu \tau}\sum_{C\gamma} \Biggl[ - \frac{i\hbar P_{C\gamma}}{M_C}\Bigl((\Gamma_{\mu \nu}^{C\gamma})^{[R_{A\alpha}]}(D^{\tau \tau}_{\mu \nu})^{[R_{B\beta}] } + (\Gamma_{\mu \nu}^{C\gamma})^{[R_{A\alpha}R_{B\beta}]}D^{\tau \tau}_{\mu \nu}\Bigr) \\
\nonumber
& & -  \frac{\hbar^2}{2 M_C}\Bigl((\tilde{\Gamma}^{C\gamma}S\tilde{\Gamma}^{C\gamma})_{\mu\nu}^{[R_{A\alpha}]}(D^{\tau \tau}_{\mu \nu})^{[R_{B\beta}]} + (\tilde{\Gamma}^{C\gamma}S\tilde{\Gamma}^{C\gamma})_{\mu\nu}^{[R_{A\alpha}R_{B\beta}]}D^{\tau \tau}_{\mu \nu}\Bigr)\Biggr]\\
& & + \left(\frac{\partial^2 E}{\partial R_{A\alpha} \partial R_{B\beta}}\right)_{eq}
\end{eqnarray}
 Here $\left(\frac{\partial^2 E}{\partial R_{A\alpha} \partial R_{B\beta}}\right)_{eq}$ is the regular BO Hessian expression. 
 
 Note that, for systems with formal time-reversal symmetry and without degeneracy or near degeneracy in the ground state, we expect that the minimum energy of the phase space Hamiltonian will occur at $\bP = 0$.  In such a case, the electronic wavefunction will be real-valued. Furthermore, note that any real function must also have real derivatives, thus, the density derivatives must also be real and $D_{\mu\nu}^{[R]}$ must be symmetric (since $D_{\mu\nu}^{[R]} = \sum_i C_{\mu i}^{[R]}C_{\nu i} + C_{\mu i}C_{\nu i}^R = D_{\nu\mu}^{[R]}$). At the same time, it is clear that $\Gamma$ and $\Gamma^{[R]}$ are anti-symmetric (as seen in Appendix \ref{appendix:gammaredefine}). 
Thus, according to Eq. \ref{eq:dEdPdR}, we find that $\frac{\partial^2 E_{PS}}{\partial P_{A\alpha} \partial R_{B\beta}}$ should always be zero. 

As a sidenote, exploring degenerate ground states when these simplifications would no longer hold will be extremely interesting in the future.

\section{Magnetic Field Perturbation Equivalence}\label{appendix:MFP}

In this section of the Appendix, for the sake of completeness, we review MFP for the reader and show why a phase-space Hamiltonian approach to a VCD spectra is equivalent to MFP\cite{Stephens1985} -- if one were to include the entire derivative coupling (see Eq. \ref{eq:decompose_d}), instead of $\Gamma$ in the equations for $\pp{\bmm}{\bP}$. Let us begin by considering the perturbative expansion of our BO wavefunction using the derivative coupling as our perturbation, $H' =-i\hbar \frac{\bm{P}\cdot \grad}{M} $. According to pertrubation theory, the perturbed electronic ground state state can be written down as a sum over all other electronic states $n$:

\begin{equation}
\label{eq:psi_pet}
\begin{aligned}
\ket{\Phi_{G}} &=\ket{\Phi_{G}}_{eq} + \sum_{A \alpha}\Biggl(\frac{-i\hbar}{M_A}\sum_{n\neq G}   \frac{\bra{\Phi_{n}} \grad \ket{\Phi_{G}} }{(E_G - E_n)} \ket{\Phi_{n}}  \Biggr)_{eq}P_{A\alpha}\\
&=\ket{\Phi_{G}}_{eq} + \sum_{A \alpha}\Biggl(\frac{-i\hbar}{M_A}\sum_{n\neq G}   \frac{d^{A \alpha}_{nG}}{(E_G - E_n)} \ket{\Phi_{n}}  \Biggr)_{eq}P_{A\alpha}
\end{aligned}
\end{equation}

\noindent Viewed as a  Taylor expansion, Eq. \ref{eq:psi_pet} implies that  
\begin{equation}
\begin{aligned}
\ket{\frac{\partial \Phi_{G}}{\partial P_{A \alpha}}}_{eq}  &=  \frac{-i\hbar}{M_A}  \sum_{n\neq G} \frac{d^{A \alpha}_{nG}}{(E_G - E_n)} \ket{\Phi_{n}}  \\
\end{aligned}
\end{equation}

\begin{comment}
\begin{equation}
\begin{aligned}
\ket{\frac{\partial \Phi_{G} }{\partial P_{A \alpha}}} &=  \sum_{n\neq G}  \frac{\bra{n} \hat{d}^{A \alpha} \ket{\Phi_{G}^0} }{M_A(E_G - E_n)} \ket{n} \\
&=  \sum_{n\neq G}  \frac{\bra{n} \frac{\partial H}{\partial R_{A\alpha}} \ket{\Phi_{G}^0} }{M_A(E_G - E_n)^2} \ket{n}
\end{aligned}
\end{equation}
\end{comment}

Next, if we insert our pertubative wavefunction into Eq. \ref{eq:L_pet} and keep the first order terms (keeping in mind that the zeroth order terms vanish)
%Next, we can insert our pertubative wavefunction into the electronic part of the magnetic transition dipole moment and, keeping the first order terms, we find .%

\begin{equation}
\begin{aligned}
\label{eq:dPdmuSOS}
%\bra{\Phi_{G}} \hat{L}_e \ket{\Phi_{G}} 
\left(\frac{\partial}{\partial P_{A\alpha}} \bra{\Phi_{G}} \bm{\hat{\mathcal{m}}}^e \ket{\Phi_{G}}\right)_{eq} &=
 \frac{e}{2m_e} \left(  \left<\frac{\partial \Phi_{G} }{\partial P_{A \alpha}}\middle| \bm{\hat{L}^e} \middle| \Phi_{G}\right>+ \left<\Phi_{G}\middle| \bm{\hat{L}^e} \middle|\frac{\partial \Phi_{G} }{\partial P_{A \alpha}}\right> \right)_{eq}\\
&=  \frac{-i\hbar e}{2M_Am_e} \sum_{n\neq G} \left(  \frac{d^{A \alpha}_{Gn}} {(E_G - E_n)} \bra{\Phi_{n}} \bm{\hat{L}^e} \ket{\Phi_{G}} + \bra{\Phi_{G}} \bm{\hat{L}^e} \ket{\Phi_{n}} \frac{d^{A \alpha}_{nG}}{(E_G - E_n)}\right)_{eq}\\
&= \frac{-i\hbar e}{2M_Am_e} \sum_{n\neq G} \left(  \frac{\bra{\Phi_{G}} \frac{\partial \hH}{\partial R_{A \alpha}} \ket{\Phi_{n}}} {(E_G - E_n)^2} \bra{\Phi_{n}} \bm{\hat{L}^e} \ket{\Phi_{G}} - \bra{\Phi_{G}} \bm{\hat{L}^e} \ket{\Phi_{n}} \frac{\bra{\Phi_{n}} \frac{\partial \hH}{\partial R_{A \alpha}} \ket{\Phi_{G}}}{(E_G - E_n)^2}\right)_{eq}\\
&= \frac{-i\hbar e}{M_Am_e} \left(\sum_{n\neq G} \frac{ \bra{\Phi_{G}} \frac{\partial \hH}{\partial R_{A \alpha}} \ket{\Phi_{n}} \bra{\Phi_{n}} \bm{\hat{L}^e} \ket{\Phi_{G}}} {(E_G - E_n)^2}\right)_{eq} 
\end{aligned}
\end{equation}

\noindent where Hellman-Feynman was invoked in the third step, and in the fourth step we used the fact that $\bm{\hat{L}^e}$ is Hermitian and purely imaginary. 

Next we consider the Hamiltonian under some constant perturbative
 magnetic field. This perturbation can be written as

 $$ H' =  -\bm{\hat{\mathcal{m}}}^e_{} \cdot \bm{B} = \frac{e}{2m_e} \bm{\hat{L}^e}\cdot \bm{B}$$

%Under this perturbation $\Phi_{G}$ can be Taylor expanded as

%\begin{equation}
%\begin{aligned}
%    \ket{\Phi_{G}} = \ket{\Phi_{G}}_{eq} + \sum_\beta 
%\ket{\frac{\partial \Phi_{G}}{\partial B_\beta}}_{eq}  B_\beta
%\end{aligned}
%\end{equation}

 \noindent According to first-order Rayleigh perturbation theory, the first order wavefunction $\ket{\Phi_{G}}$ becomes:

\begin{equation}
\label{eq:Phi_Bpet}
\ket{\Phi_{G}} = \ket{\Phi_{G}}_{eq} + \frac{e}{2m_e} \left( \sum_{n\neq G}  \frac{\bra{\Phi_{n}} \bm{\hat{L}^e} \ket{\Phi_{G}} }{(E_G - E_n)} \ket{\Phi_{n}}\right)_{eq} \cdot \bm{B}
\end{equation}

Viewed as a Taylor expansion, Eq. \ref{eq:Phi_Bpet} implies that 

\begin{equation}\ket{\frac{\partial \Phi_{G}}{\partial B_\beta}}_{eq} =  \frac{e}{2m_e} \left( \sum_{n\neq G}  \frac{\bra{\Phi_{n}} \hat{L}^e_\beta \ket{\Phi_{G}} }{(E_G - E_n)} \ket{\Phi_{n}}\right)_{eq}
\end{equation}

Next, if we  perturb the Hamiltonian by a nuclear coordinate perturbation, $ \hH' =  \sum_{A \alpha}(\pp{\hH}{R_{A\alpha}})_{eq} \cdot (R_{A\alpha}-R^{eq}_{A\alpha})$, we find one more relevant expression:

\begin{equation}
\bra{\Phi_{G}} = \bra{\Phi_{G}}_{eq} + \sum_{A \alpha} \left(\sum_{m\neq G}  \frac{\bra{\Phi_{G}} \frac{\partial \hH}{\partial R_{A\alpha}} \ket{\Phi_{m}} }{(E_G - E_m)} \bra{\Phi_{m}} \right)_{eq}\cdot (R_{A\alpha}-R_{A\alpha}^{eq})
\end{equation}
or in other words,
\begin{equation}
\label{eq:Phi_Rpet}
\bra{\frac{\partial \Phi_{G}}{\partial R_{A\alpha}}}_{eq} = \left(\sum_{m\neq G}   \frac{\bra{\Phi_{G}} \frac{\partial \hH}{\partial R_{A\alpha}} \ket{\Phi_{m}} }{(E_G - E_m)} \bra{\Phi_{m}}\right)_{eq}
\end{equation}

\noindent Taking the inner product of \ref{eq:Phi_Rpet} and \ref{eq:Phi_Bpet}, we find:

\begin{equation}
\begin{aligned}
\label{eq:MFPSOS}
\bra{\frac{\partial{\Phi_{G}}}{\partial B_\beta}}\ket{\frac{\partial{\Phi_{G}}}{\partial R_{A\alpha}}}_{eq} &= \frac{e}{2m_e} \left(\sum_{n,m \neq G}  \frac{\bra{\Phi_{G}} \hat{L}^e_\beta \ket{\Phi_{m}} }{E_G - E_m} \frac{\bra{\Phi_{n}} \frac{\partial H}{\partial R_{A\alpha}} \ket{\Phi_{G}} }{E_G - E_n} \bra{\Phi_{m}}\ket{\Phi_{n}} \right)_{eq}\\
&= \frac{e}{2m_e} \left(\sum_{n\neq G}  \frac{ \bra{\Phi_{G}} \frac{\partial H}{\partial R_{A\alpha}} \ket{\Phi_{n}} \bra{\Phi_{n}} \hat{L}^e_\beta \ket{\Phi_{G}}} {(E_G - E_n)^2}\right)_{eq}\\
%&= \frac{M_A}{2} \left( \frac{\partial}{\partial P_{A\alpha}} \bra{\Phi_{G}} \hat{L}^e_\beta \ket{\Phi_{G}}\right)_{eq}
\end{aligned}
\end{equation}

From Eqs. \ref{eq:dPdmuSOS} and \ref{eq:MFPSOS}, it follows that 

\begin{eqnarray}
\label{eq:dmdpMFPequal}
    \left(\frac{\partial}{\partial P_{A\alpha}} \bra{\Phi_{G}} \hat{\mathcal{m}}^e_\beta \ket{\Phi_{G}}\right)_{eq} = \frac{2i\hbar}{M_A}\bra{\frac{\partial{\Phi_{G}}}{\partial B_\beta}}\ket{\frac{\partial{\Phi_{G}}}{\partial R_{A\alpha}}}_{eq}
\end{eqnarray}
which is the basis for the MFP VCD expression.
\cite{Stephens1985}

\section{\texorpdfstring{On the Calculation of $\bm{\Tilde{\Gamma}}$ and a Proof that the Kinetic Energy (Eq. \ref{eq:maybe}) is Positive Definite}{On the Calculation of Gamma and a Proof that the Kinetic Energy (Eq. \ref{eq:maybe}) is Positive Definite}}\label{appendix:posdef}

%\section{On the Calculation of $\Tilde{\Gamma}$ and a Proof that the Kinetic Energy (Eq. \ref{eq:maybe}) is Positive Definite}\label{appendix:posdef}

In this section, we will begin by showing how to evaluate $\tilde{\bGamma}$ in Eq. \ref{eq:Gtildedef} above. Starting from Eq. \ref{eq:Gtildedef}, we compute the unitary transformation which diagonalizes the overlap matrix S ($U^{\dagger}S U = s$) and

\begin{eqnarray}
\nonumber
& \frac{1}{2}(s_\lambda (U^{\dagger} \Tilde{\bGamma}U)_{\lambda\sigma} + (U^{\dagger} \Tilde{\bGamma}U)_{\lambda\sigma} s_\sigma) = (U^{\dagger} \bGamma U)_{\lambda\sigma} \\
\Rightarrow & \tilde{\bGamma}^A_{\mu\nu} = \sum_{\sigma\lambda}2U_{\mu\sigma} ((s_\lambda + s_\sigma)^{-1}  (U^\dagger \bGamma^A U)_{\sigma\lambda})(U^\dagger)_{\lambda\nu} 
\end{eqnarray}

%don't have U^{\dagger} with indices, not clear

Next, let us show that Eq. \ref{eq:eps_stable} is positive-definite in momentum $\bP$. For simplicity, let us suppose we have a single nuclear degree of freedom. According to Eq. \ref{eq:eps_stable},
the nuclear kinetic energy is of the following form:

\begin{eqnarray}
\label{eq:<Tnuc>}
    \langle \hat{T}_{nuc} \rangle &=& \frac{1}{2M}\left(P^2 - i \hbar \mbox{Tr}\left((\tilde{\Gamma} S + S \tilde{\Gamma}) D \right)P - \hbar^2 \mbox{Tr}\left((\tilde{\Gamma} S \tilde{\Gamma}) D \right) \right)
\end{eqnarray}
Note that, in this expression, $\Gamma, S, \text{ and } D$ are all matrices. Now, let $D_S  = S^{1/2}D S^{1/2}$, and  $\Gamma_S = i \hbar S^{1/2} \tilde{\Gamma} S^{-1/2} $.
Note that $\Gamma_S^{\dagger} = i \hbar S^{-1/2} \tilde{\Gamma} S^{1/2} $ and that $D_S^2 = D_S$. Under this transformation, $\langle T_{nuc} \rangle$ can be rewritten as

\begin{equation}
\begin{aligned}
    \langle \hat{T}_{nuc} \rangle &= \frac{1}{2M}\Biggl[P^2 - \mbox{Tr}\left((\Gamma_S^{\dagger} + \Gamma_S) D_S \right)P + \mbox{Tr}(\Gamma_S^{\dagger} \Gamma_S D_S)\Biggr] \\ 
     & = \frac{1}{2M} \Biggl[ P^2 - \mbox{Tr}\left((\Gamma_S^{\dagger} + \Gamma_S) D_S \right)P + \mbox{Tr}(\Gamma_S^{\dagger} D_S)   \mbox{Tr}(\Gamma_S D_S)  \\
     &  + \mbox{Tr}(\Gamma_S^{\dagger} \Gamma_S D_S) - \mbox{Tr}(\Gamma_S^{\dagger} D_S)   \mbox{Tr}(\Gamma_S D_S)\Biggr] \\ 
     &=  \frac{1}{2M}\Biggl[\Bigl(P - \mbox{Tr}(\Gamma_S D_S)\Bigr) \Bigl(P - \mbox{Tr}(\Gamma_S^{\dagger} D_S)\Bigr)  \\
     &  + \mbox{Tr}(\Gamma_S^{\dagger} \Gamma_S D_S) - \mbox{Tr}(\Gamma_S^{\dagger} D_S)   \mbox{Tr}(\Gamma_S D_S)  \Biggr]
     \\
     &= \frac{1}{2M}\Biggl[\Bigl(P - \mbox{Tr}(\Gamma_S D_S)\Bigr) \Bigl(P - \mbox{Tr}(\Gamma_S^{\dagger} D_S)\Bigr) \\
     & + \mbox{Tr}\Biggl(\Bigl((D_S\Gamma_S)^{\dagger} -  \mbox{Tr}(\Gamma_S^{\dagger} D_S)  \Bigr)  \Bigl(D_S\Gamma_S -  \mbox{Tr}(\Gamma_S D_S)  \Bigr) \Biggr)\Biggr]
\end{aligned}
\end{equation}

Since both terms in this expression are squared scalars, the entire quantity must be positive definite.

\section{\texorpdfstring{Explicit Definitions of $\bGamma$ from Ref. \cite{tian24gamma}}{Explicit Definitions of Gamma from Ref. tian24gamma) is Positive Definite}}\label{appendix:gammaredefine}

%\section{Explicit Definitions of $\Gamma$ from Ref. \cite{tian24gamma}\label{appendix:gammaredefine}}

For the sake of concreteness, we will here repeat the exact definitions of $\bGamma'$ and $\bGamma''$ from Ref. \citenum{tian24gamma}. First, as far $\bGamma'$ is concerned, we set  
\begin{align}
\Gamma'^{A \alpha}_{\mu \nu}  = \frac{1}{2i\hbar} p^{\alpha}_{\mu \nu} \left( \delta_{BA} + \delta_{CA}\right)\label{eq:etf}
\end{align}
just as in Eq. \ref{eq:detf} above.
Here and below, $\mu$ indexes an orbital centered on atom $B$, $\nu$ indexes an orbital centered on atom $C$, and $p^{\alpha}_{\mu \nu}$ is the $\alpha$-component of the electronic momentum. Intuitively, the electronic momentum operator emerges because we must take into account the fact that any nuclear displacement moves the electrons as well. 

As far as the definition of $\bGamma''$ is concerned, we set
\begin{align}
    \bm{\Gamma}''^{A}_{\mu\nu} &= \zeta_{\mu\nu}^A\bm{X}_A \times \left(\bm{K}_{\mu\nu}^{-1}\bm{J}^A_{\mu\nu}\right)\label{eq:gamma_v2_scale}\\
    \bm{K}_{\mu\nu}&=-\sum_{A}\zeta_{\mu\nu}^A\left(\bm{X}_A^\top\bm{X}_A\right)\bm{I} + \sum_A\zeta_{\mu\nu}^A\bm{X}_A\bm{X}_A^\top\label{eq:K_scale}\\
    \zeta^A_{\mu\nu} &= \exp\left(-w \frac{2|(\bm{X}_A-\bm{X}_B)|^2 |(\bm{X}_A-\bm{X}_C)|^2}{|(\bm{X}_A-\bm{X}_B)|^2 + |(\bm{X}_A-\bm{X}_C)|^2}\right)\label{eq:zeta}\\
    \bm{J}_{\mu\nu} &= \frac{1}{i\hbar}\left<\chi_\mu \middle |\frac{1}{2}\left(\hat{\bm{l}}^{(B)}+\hat{\bm{l}}^{(C)}\right) + \hat{\bm{s}} \middle|\chi_\nu \right>\label{eq:J}
\end{align}

Here, $\bm{J}_{\mu\nu}$ is the atomic orbital centered electronic angular momentum, $\hat{\bm{l}}^{(A)}$ is the electronic momentum around atom A, $\zeta^A_{\mu\nu}$ is a weighting factor to maintain semi-locality of $\bGamma''$ (we set the parameter $w=0.3$),\cite{tian24gamma} $\bm{I}$ is the $3 \times 3$ identity matrix, and the matrix $\bK_{\mu\nu}$ is effectively the negative of a locality weighted massless moment of inertia in the vicinity of the $\chi_{\mu}$ and $\chi_{\nu}$ orbitals.

Obviously, $\bm{\Gamma'}$ and $\bm{\Gamma''}$ are anti-Hermitian. Moreover, if there is no spin-orbit coupling and we ignore the spin ``$\hat{\bm{s}}$'' term in Eq. \ref{eq:J}, these matrices are also purely real and anti-symmetric.

\section{Data Tables}\label{appendix:tables}

\begin{table}[H]
\caption{Experimental rotational strength of R-$d_2$-oxirane as compared with {\em ab initio} calculations using the  $\Gamma$ couplings above or MFP. Experimental data\cite{expvcdmeoxi} in $C_2Cl_4$ solution for the C-H and C-D stretching modes and in $CS_2$ solution otherwise. \label{tbl:oxi_exp} }
\begin{tabular}{cccccccc}
\hline \hline
\multicolumn{2}{l}{\textbf{Experimental}}              & \multicolumn{3}{l}{\textbf{MFP/aug-cc-pvqz}} & \multicolumn{3}{l}{$\mathbf{\Gamma}$\textbf{/aug-cc-pvqz}} \\ \hline
Freq	&	$\mathcal{R}_{expt}$	&	Freq	&	$\mathcal{R}$	&	Sign	&	Freq	&	$\mathcal{R}$	&	Sign	\\	\hline
673	&		&	714	&	1.2	&		&	719	&	1.1	&		\\	
754	&		&	835	&	16.3	&		&	839.0	&	11	&		\\	
917	&	+	&	956	&	0.3	&		&	959.0	&	-2.6	&	X	\\	
885	&	5	&	978	&	5.5	&		&	984.0	&	3.5	&		\\	
914	&	-6	&	1031	&	0.3	&	X	&	1034.0	&	-4.6	&		\\	
961	&	-29	&	1080	&	-46.7	&		&	1083.0	&	-33.9	&		\\	
1106	&	11	&	1251	&	9.1	&		&	1247.0	&	18.6	&		\\	
1112	&	-5	&	1267	&	9.5	&	X	&	1263.0	&	7.1	&	X	\\	
1235	&	24	&	1364	&	13.9	&		&	1367.0	&	11.9	&		\\	
1339	&	-3	&	1479	&	-2.1	&		&	1476.0	&	-0.8	&		\\	
1397	&	-15	&	1560	&	-11.4	&		&	1560.0	&	-9.8	&		\\	
2240	&	-10	&	2412	&	-16.4	&		&	2403.0	&	-8.8	&		\\	
2254	&	12	&	2428	&	15.8	&		&	2419.0	&	12.4	&		\\	
3015	&	-9	&	3291	&	-28.2	&		&	3256.0	&	-20.9	&		\\	
3028	&	11	&	3295	&	33.6	&		&	3260.0	&	18.3	&		\\	
\hline\hline
\end{tabular}
\end{table}

\begin{table}[H]
\caption{Experimental rotational strength of R-$d_2$-cyclopropane\cite{expvcdcycp} as compared with {\em ab initio} calculations using the  $\Gamma$ couplings above or MFP \label{tbl:cycP_exp}.}
\begin{tabular}{cccccccc}
\hline \hline
\multicolumn{2}{l}{\textbf{Experimental}}              & \multicolumn{3}{l}{\textbf{MFP/aug-cc-pvqz}} & \multicolumn{3}{l}{$\mathbf{\Gamma}$\textbf{/aug-cc-pvqz}} \\ \hline
Freq	&	$\mathcal{R}_{expt}$	&	Freq	&	$\mathcal{R}$	&	Sign	&	Freq	&	$\mathcal{R}$	&	Sign	\\	\hline
618	&		&	668	&	0.4	&		&	679	&	0.5	&		\\	
632	&		&	683	&	-0.8	&		&	692	&	-2.6	&		\\	
736	&		&	791	&	2.5	&		&	799	&	1.1	&		\\	
786	&		&	850	&	-9.7	&		&	858	&	0.7	&		\\	
857	&		&	926	&	4.3	&		&	933	&	-0.3	&		\\	
909	&		&	1013	&	6.0	&		&	1020	&	1.8	&		\\	
942	&	5.5	&	1076	&	0.4	&		&	1080	&	0.5	&		\\	
1037	&	-15.9	&	1186	&	-5.6	&		&	1186	&	-3.1	&		\\	
1052	&	28.2	&	1162	&	15.0	&		&	1165	&	11.2	&		\\	
1087	&	-13.3	&	1221	&	-0.9	&		&	1216	&	-1.1	&		\\	
1134	&	-42.6	&	1262	&	-17.2	&		&	1265	&	-11.7	&		\\	
1180	&	3	&	1293	&	3.5	&		&	1298	&	3.0	&		\\	
1290	&	10.2	&	1435	&	6.3	&		&	1435	&	4.9	&		\\	
1338	&	-7.7	&	1491	&	-4.7	&		&	1491	&	-3.6	&		\\	
	&		&	1620	&	0.6	&		&	1616	&	0.6	&		\\	
2257	&	3.3	&	2426	&	10.1	&		&	2421	&	6.1	&		\\	
2270	&	-3.7	&	2438	&	-10.5	&		&	2431	&	-6.6	&		\\	
3012	&	1.4	&	3266	&	6.8	&		&	3237	&	3.7	&		\\	
3035	&	4.4	&	3306	&	21.6	&		&	3276	&	11.1	&		\\	
3041	&	-9.9	&	3308	&	-38.7	&		&	3279	&	-20.8	&		\\	
3076	&	4.8	&	3345	&	10.6	&		&	3315	&	6.1	&		\\	
\hline\hline
\end{tabular}
\end{table}

\begin{table}
\caption{Experimental\cite{expvcdmeoxi} rotational strength of S-propylene-oxide as compared with {\em ab initio} calculations using the  $\Gamma$ couplings above or MFP \label{tbl:meoxi_exp}. Mode assignments based on Ref. \citenum{assignmodemeoxi}.}
\begin{tabular}{cccccccc}
\hline \hline
\multicolumn{2}{l}{\textbf{Experimental}}              & \multicolumn{3}{l}{\textbf{MFP/aug-cc-pvqz}} & \multicolumn{3}{l}{$\mathbf{\Gamma}$\textbf{/aug-cc-pvqz}} \\ \hline
Freq	&	$\mathcal{R}_{expt}$	&	Freq	&	$\mathcal{R}$	&	Sign	&	Freq	&	$\mathcal{R}$	&	Sign	\\	\hline
200	&		&	227	&	-3.5	&		&	219	&	-1.3	&		\\	
373	&		&	397	&	12.7	&		&	396	&	15.1	&		\\	
414	&		&	442	&	7.3	&		&	441	&	5.0	&		\\	
748	&	-21.6	&	851	&	-6.8	&		&	847	&	-6.5	&		\\	
828	&		&	940	&	-7.9	&		&	937	&	2.5	&		\\	
893	&	-42.8	&	986	&	-25.5	&		&	979	&	-18.5	&		\\	
951	&	71.9	&	1070	&	25.4	&		&	1066	&	18.3	&		\\	
1022	&	-6.7	&	1139	&	-12.5	&		&	1132	&	-7.7	&		\\	
1130	&	8.1	&	1267	&	10.0	&		&	1261	&	3.1	&		\\	
1143	&	-15.9	&	1296	&	2.9	&	X	&	1287	&	0.1	&	X	\\	
1165	&	21.4	&	1310	&	7.3	&		&	1301	&	10.7	&		\\	
1103	&	-0.4	&	1238	&	10.2	&	X	&	1232	&	-4.7	&		\\	
1262	&	15.6	&	1406	&	8.7	&		&	1399	&	1.5	&		\\	
1368	&	-3.3	&	1526	&	-2.9	&		&	1515	&	-3.3	&		\\	
1406	&	-8.4	&	1574	&	-11.0	&		&	1565	&	-10.1	&		\\	
1444	&		&	1600	&	-3.1	&		&	1591	&	-1.3	&		\\	
1456	&		&	1614	&	-0.2	&		&	1605	&	2.4	&		\\	
1500	&	-4	&	1672	&	-8.0	&		&	1662	&	2.5	&	X	\\	
2928	&		&	3162	&	-1.1	&		&	3127	&	0.7	&		\\	
2970	&		&	3217	&	7.0	&		&	3186	&	0.3	&		\\	
2970	&		&	3231	&	-10.8	&		&	3200	&	-11.6	&		\\	
2995	&		&	3242	&	14.9	&		&	3208	&	17.4	&		\\	
2995	&		&	3261	&	-10.8	&		&	3228	&	-10.5	&		\\	
3046	&	3.5	&	3325	&	5.5	&		&	3290	&	5.8	&		\\	
																		
\hline\hline
\end{tabular}
\end{table}

\typeout{}
 \pagebreak

\bibliography{reference}

\providecommand{\latin}[1]{#1}
\makeatletter
\providecommand{\doi}
  {\begingroup\let\do\@makeother\dospecials
  \catcode`\{=1 \catcode`\}=2 \doi@aux}
\providecommand{\doi@aux}[1]{\endgroup\texttt{#1}}
\makeatother
\providecommand*\mcitethebibliography{\thebibliography}
\csname @ifundefined\endcsname{endmcitethebibliography}  {\let\endmcitethebibliography\endthebibliography}{}
\begin{mcitethebibliography}{65}
\providecommand*\natexlab[1]{#1}
\providecommand*\mciteSetBstSublistMode[1]{}
\providecommand*\mciteSetBstMaxWidthForm[2]{}
\providecommand*\mciteBstWouldAddEndPuncttrue
  {\def\EndOfBibitem{\unskip.}}
\providecommand*\mciteBstWouldAddEndPunctfalse
  {\let\EndOfBibitem\relax}
\providecommand*\mciteSetBstMidEndSepPunct[3]{}
\providecommand*\mciteSetBstSublistLabelBeginEnd[3]{}
\providecommand*\EndOfBibitem{}
\mciteSetBstSublistMode{f}
\mciteSetBstMaxWidthForm{subitem}{(\alph{mcitesubitemcount})}
\mciteSetBstSublistLabelBeginEnd
  {\mcitemaxwidthsubitemform\space}
  {\relax}
  {\relax}

\bibitem[Gal(2008)]{Pasteur}
Gal,~J. The discovery of biological enantioselectivity: Louis Pasteur and the fermentation of tartaric acid, 1857—A review and analysis 150 yr later. \emph{Chirality} \textbf{2008}, \emph{20}, 5--19\relax
\mciteBstWouldAddEndPuncttrue
\mciteSetBstMidEndSepPunct{\mcitedefaultmidpunct}
{\mcitedefaultendpunct}{\mcitedefaultseppunct}\relax
\EndOfBibitem
\bibitem[List \latin{et~al.}(2000)List, Lerner, and Barbas]{NobelList}
List,~B.; Lerner,~R.~A.; Barbas,~C.~F. Proline-Catalyzed Direct Asymmetric Aldol Reactions. \emph{Journal of the American Chemical Society} \textbf{2000}, \emph{122}, 2395--2396\relax
\mciteBstWouldAddEndPuncttrue
\mciteSetBstMidEndSepPunct{\mcitedefaultmidpunct}
{\mcitedefaultendpunct}{\mcitedefaultseppunct}\relax
\EndOfBibitem
\bibitem[Northrup and MacMillan(2002)Northrup, and MacMillan]{NobelMacMillan}
Northrup,~A.~B.; MacMillan,~D.~W. The first direct and enantioselective cross-aldol reaction of aldehydes. \emph{Journal of the American Chemical Society} \textbf{2002}, \emph{124}, 6798--6799\relax
\mciteBstWouldAddEndPuncttrue
\mciteSetBstMidEndSepPunct{\mcitedefaultmidpunct}
{\mcitedefaultendpunct}{\mcitedefaultseppunct}\relax
\EndOfBibitem
\bibitem[Naaman and Waldeck(2012)Naaman, and Waldeck]{Naaman2012}
Naaman,~R.; Waldeck,~D.~H. Chiral-induced spin selectivity effect. \emph{The journal of physical chemistry letters} \textbf{2012}, \emph{3}, 2178--2187, Publisher: ACS Publications\relax
\mciteBstWouldAddEndPuncttrue
\mciteSetBstMidEndSepPunct{\mcitedefaultmidpunct}
{\mcitedefaultendpunct}{\mcitedefaultseppunct}\relax
\EndOfBibitem
\bibitem[Naaman \latin{et~al.}(2019)Naaman, Paltiel, and Waldeck]{Naaman2019}
Naaman,~R.; Paltiel,~Y.; Waldeck,~D.~H. Chiral molecules and the electron spin. \emph{Nature Reviews Chemistry} \textbf{2019}, \emph{3}, 250--260, Publisher: Nature Publishing Group UK London\relax
\mciteBstWouldAddEndPuncttrue
\mciteSetBstMidEndSepPunct{\mcitedefaultmidpunct}
{\mcitedefaultendpunct}{\mcitedefaultseppunct}\relax
\EndOfBibitem
\bibitem[Evers \latin{et~al.}(2022)Evers, Aharony, Bar-Gill, Entin-Wohlman, Hedegård, Hod, Jelinek, Kamieniarz, Lemeshko, Michaeli, Mujica, Naaman, Paltiel, Refaely-Abramson, Tal, Thijssen, Thoss, van Ruitenbeek, Venkataraman, Waldeck, Yan, and Kronik]{Evers2022}
Evers,~F.; Aharony,~A.; Bar-Gill,~N.; Entin-Wohlman,~O.; Hedegård,~P.; Hod,~O.; Jelinek,~P.; Kamieniarz,~G.; Lemeshko,~M.; Michaeli,~K.; Mujica,~V.; Naaman,~R.; Paltiel,~Y.; Refaely-Abramson,~S.; Tal,~O.; Thijssen,~J.; Thoss,~M.; van Ruitenbeek,~J.~M.; Venkataraman,~L.; Waldeck,~D.~H.; Yan,~B.; Kronik,~L. Theory of Chirality Induced Spin Selectivity: Progress and Challenges. \emph{Adv. Mater.} \textbf{2022}, \emph{34}, 2106629\relax
\mciteBstWouldAddEndPuncttrue
\mciteSetBstMidEndSepPunct{\mcitedefaultmidpunct}
{\mcitedefaultendpunct}{\mcitedefaultseppunct}\relax
\EndOfBibitem
\bibitem[Zhang and Niu(2015)Zhang, and Niu]{Zhang2015}
Zhang,~L.; Niu,~Q. Chiral phonons at high-symmetry points in monolayer hexagonal lattices. \emph{Physical review letters} \textbf{2015}, \emph{115}, 115502\relax
\mciteBstWouldAddEndPuncttrue
\mciteSetBstMidEndSepPunct{\mcitedefaultmidpunct}
{\mcitedefaultendpunct}{\mcitedefaultseppunct}\relax
\EndOfBibitem
\bibitem[Bistoni \latin{et~al.}(2021)Bistoni, Mauri, and Calandra]{Bistoni2021}
Bistoni,~O.; Mauri,~F.; Calandra,~M. Intrinsic vibrational angular momentum from nonadiabatic effects in noncollinear magnetic molecules. \emph{Physical Review Letters} \textbf{2021}, \emph{126}, 225703\relax
\mciteBstWouldAddEndPuncttrue
\mciteSetBstMidEndSepPunct{\mcitedefaultmidpunct}
{\mcitedefaultendpunct}{\mcitedefaultseppunct}\relax
\EndOfBibitem
\bibitem[Das \latin{et~al.}(2022)Das, Tassinari, Naaman, and Fransson]{Das2022}
Das,~T.~K.; Tassinari,~F.; Naaman,~R.; Fransson,~J. Temperature-dependent chiral-induced spin selectivity effect: Experiments and theory. \emph{The Journal of Physical Chemistry C} \textbf{2022}, \emph{126}, 3257--3264\relax
\mciteBstWouldAddEndPuncttrue
\mciteSetBstMidEndSepPunct{\mcitedefaultmidpunct}
{\mcitedefaultendpunct}{\mcitedefaultseppunct}\relax
\EndOfBibitem
\bibitem[Fransson(2023)]{Fransson2023}
Fransson,~J. Chiral phonon induced spin polarization. \emph{Physical Review Research} \textbf{2023}, \emph{5}, L022039, Publisher: APS\relax
\mciteBstWouldAddEndPuncttrue
\mciteSetBstMidEndSepPunct{\mcitedefaultmidpunct}
{\mcitedefaultendpunct}{\mcitedefaultseppunct}\relax
\EndOfBibitem
\bibitem[Kim \latin{et~al.}(2023)Kim, Vetter, Yan, Yang, Wang, Sun, Yang, Comstock, Li, Zhou, \latin{et~al.} others]{Kim2023}
Kim,~K.; Vetter,~E.; Yan,~L.; Yang,~C.; Wang,~Z.; Sun,~R.; Yang,~Y.; Comstock,~A.~H.; Li,~X.; Zhou,~J.; others Chiral-phonon-activated spin Seebeck effect. \emph{Nature Materials} \textbf{2023}, \emph{22}, 322--328\relax
\mciteBstWouldAddEndPuncttrue
\mciteSetBstMidEndSepPunct{\mcitedefaultmidpunct}
{\mcitedefaultendpunct}{\mcitedefaultseppunct}\relax
\EndOfBibitem
\bibitem[Berova \latin{et~al.}(2000)Berova, Nakanishi, and Woody]{Berova2000}
Berova,~N.; Nakanishi,~K.; Woody,~R.~W. \emph{Circular dichroism: principles and applications}; John Wiley \& Sons, 2000\relax
\mciteBstWouldAddEndPuncttrue
\mciteSetBstMidEndSepPunct{\mcitedefaultmidpunct}
{\mcitedefaultendpunct}{\mcitedefaultseppunct}\relax
\EndOfBibitem
\bibitem[Warnke and Furche(2012)Warnke, and Furche]{Warnke2012}
Warnke,~I.; Furche,~F. Circular dichroism: electronic. \emph{Wiley Interdisciplinary Reviews: Computational Molecular Science} \textbf{2012}, \emph{2}, 150--166\relax
\mciteBstWouldAddEndPuncttrue
\mciteSetBstMidEndSepPunct{\mcitedefaultmidpunct}
{\mcitedefaultendpunct}{\mcitedefaultseppunct}\relax
\EndOfBibitem
\bibitem[Nafie \latin{et~al.}(1976)Nafie, Keiderling, and Stephens]{Nafie1976}
Nafie,~L.~A.; Keiderling,~T.; Stephens,~P. Vibrational circular dichroism. \emph{Journal of the American Chemical Society} \textbf{1976}, \emph{98}, 2715--2723\relax
\mciteBstWouldAddEndPuncttrue
\mciteSetBstMidEndSepPunct{\mcitedefaultmidpunct}
{\mcitedefaultendpunct}{\mcitedefaultseppunct}\relax
\EndOfBibitem
\bibitem[Stephens(1970)]{Stephens1970}
Stephens,~P. Theory of magnetic circular dichroism. \emph{The Journal of Chemical Physics} \textbf{1970}, \emph{52}, 3489--3516\relax
\mciteBstWouldAddEndPuncttrue
\mciteSetBstMidEndSepPunct{\mcitedefaultmidpunct}
{\mcitedefaultendpunct}{\mcitedefaultseppunct}\relax
\EndOfBibitem
\bibitem[Crawford \latin{et~al.}(2007)Crawford, Tam, and Abrams]{ECDReview}
Crawford,~T.~D.; Tam,~M.~C.; Abrams,~M.~L. The current state of ab initio calculations of optical rotation and electronic circular dichroism spectra. \emph{The Journal of Physical Chemistry A} \textbf{2007}, \emph{111}, 12057--12068\relax
\mciteBstWouldAddEndPuncttrue
\mciteSetBstMidEndSepPunct{\mcitedefaultmidpunct}
{\mcitedefaultendpunct}{\mcitedefaultseppunct}\relax
\EndOfBibitem
\bibitem[Nafie(1997)]{VCDROAReviewNafie}
Nafie,~L.~A. Infrared and Raman vibrational optical activity: theoretical and experimental aspects. \emph{Annual Review of Physical Chemistry} \textbf{1997}, \emph{48}, 357--386\relax
\mciteBstWouldAddEndPuncttrue
\mciteSetBstMidEndSepPunct{\mcitedefaultmidpunct}
{\mcitedefaultendpunct}{\mcitedefaultseppunct}\relax
\EndOfBibitem
\bibitem[Stephens(1974)]{MCDReview}
Stephens,~P. Magnetic circular dichroism. \emph{Annual Review of Physical Chemistry} \textbf{1974}, \emph{25}, 201--232\relax
\mciteBstWouldAddEndPuncttrue
\mciteSetBstMidEndSepPunct{\mcitedefaultmidpunct}
{\mcitedefaultendpunct}{\mcitedefaultseppunct}\relax
\EndOfBibitem
\bibitem[Stephens(1985)]{Stephens1985}
Stephens,~P.~J. Theory of vibrational circular dichroism. \emph{The Journal of Physical Chemistry} \textbf{1985}, \emph{89}, 748--752\relax
\mciteBstWouldAddEndPuncttrue
\mciteSetBstMidEndSepPunct{\mcitedefaultmidpunct}
{\mcitedefaultendpunct}{\mcitedefaultseppunct}\relax
\EndOfBibitem
\bibitem[Nafie and Diem(1979)Nafie, and Diem]{nafie1979optical}
Nafie,~L.~A.; Diem,~M. Optical activity in vibrational transitions: vibrational circular dichroism and Raman optical activity. \emph{Accounts of Chemical Research} \textbf{1979}, \emph{12}, 296--302\relax
\mciteBstWouldAddEndPuncttrue
\mciteSetBstMidEndSepPunct{\mcitedefaultmidpunct}
{\mcitedefaultendpunct}{\mcitedefaultseppunct}\relax
\EndOfBibitem
\bibitem[Freedman and Nafie(1987)Freedman, and Nafie]{Nafie1987book}
Freedman,~T.~B.; Nafie,~L.~A. \emph{Topics in Stereochemistry}; John Wiley \& Sons, Ltd, 1987; pp 113--206\relax
\mciteBstWouldAddEndPuncttrue
\mciteSetBstMidEndSepPunct{\mcitedefaultmidpunct}
{\mcitedefaultendpunct}{\mcitedefaultseppunct}\relax
\EndOfBibitem
\bibitem[Cohan and Hameka(1966)Cohan, and Hameka]{cohan1966BO}
Cohan,~N.~V.; Hameka,~H.~F. Isotope Effects in Optical Rotation1. \emph{Journal of the American Chemical Society} \textbf{1966}, \emph{88}, 2136--2142\relax
\mciteBstWouldAddEndPuncttrue
\mciteSetBstMidEndSepPunct{\mcitedefaultmidpunct}
{\mcitedefaultendpunct}{\mcitedefaultseppunct}\relax
\EndOfBibitem
\bibitem[Walnut and Nafie(1977)Walnut, and Nafie]{Nafie1977BO}
Walnut,~T.~H.; Nafie,~L.~A. Infrared absorption and the Born--Oppenheimer approximation. II. Vibrational circular dichroism. \emph{The Journal of Chemical Physics} \textbf{1977}, \emph{67}, 1501--1510\relax
\mciteBstWouldAddEndPuncttrue
\mciteSetBstMidEndSepPunct{\mcitedefaultmidpunct}
{\mcitedefaultendpunct}{\mcitedefaultseppunct}\relax
\EndOfBibitem
\bibitem[Nafie and Freedman(1983)Nafie, and Freedman]{nafie1983SOS}
Nafie,~L.~A.; Freedman,~T.~B. Vibronic coupling theory of infrared vibrational transitions. \emph{The Journal of Chemical Physics} \textbf{1983}, \emph{78}, 7108--7116\relax
\mciteBstWouldAddEndPuncttrue
\mciteSetBstMidEndSepPunct{\mcitedefaultmidpunct}
{\mcitedefaultendpunct}{\mcitedefaultseppunct}\relax
\EndOfBibitem
\bibitem[Buckingham \latin{et~al.}(1987)Buckingham, Fowler, and Galwas]{BUCKINGHAM1987}
Buckingham,~A.; Fowler,~P.; Galwas,~P. Velocity-dependent property surfaces and the theory of vibrational circular dichroism. \emph{Chemical Physics} \textbf{1987}, \emph{112}, 1--14\relax
\mciteBstWouldAddEndPuncttrue
\mciteSetBstMidEndSepPunct{\mcitedefaultmidpunct}
{\mcitedefaultendpunct}{\mcitedefaultseppunct}\relax
\EndOfBibitem
\bibitem[Nafie(1992)]{Nafie1992}
Nafie,~L.~A. {Velocity‐gauge formalism in the theory of vibrational circular dichroism and infrared absorption}. \emph{The Journal of Chemical Physics} \textbf{1992}, \emph{96}, 5687--5702\relax
\mciteBstWouldAddEndPuncttrue
\mciteSetBstMidEndSepPunct{\mcitedefaultmidpunct}
{\mcitedefaultendpunct}{\mcitedefaultseppunct}\relax
\EndOfBibitem
\bibitem[Ditler \latin{et~al.}(2022)Ditler, Zimmermann, Kumar, and Luber]{ditler2022}
Ditler,~E.; Zimmermann,~T.; Kumar,~C.; Luber,~S. Implementation of nuclear velocity perturbation and magnetic field perturbation theory in CP2K and their application to vibrational circular dichroism. \emph{Journal of Chemical Theory and Computation} \textbf{2022}, \emph{18}, 2448--2461\relax
\mciteBstWouldAddEndPuncttrue
\mciteSetBstMidEndSepPunct{\mcitedefaultmidpunct}
{\mcitedefaultendpunct}{\mcitedefaultseppunct}\relax
\EndOfBibitem
\bibitem[Scherrer \latin{et~al.}(2013)Scherrer, Vuilleumier, and Sebastiani]{Sebastiani2013}
Scherrer,~A.; Vuilleumier,~R.; Sebastiani,~D. Nuclear Velocity Perturbation Theory of Vibrational Circular Dichroism. \emph{Journal of Chemical Theory and Computation} \textbf{2013}, \emph{9}, 5305--5312\relax
\mciteBstWouldAddEndPuncttrue
\mciteSetBstMidEndSepPunct{\mcitedefaultmidpunct}
{\mcitedefaultendpunct}{\mcitedefaultseppunct}\relax
\EndOfBibitem
\bibitem[Micha(1983)]{MichaPSSH}
Micha,~D.~A. {A self‐consistent eikonal treatment of electronic transitions in molecular collisions}. \emph{The Journal of Chemical Physics} \textbf{1983}, \emph{78}, 7138--7145\relax
\mciteBstWouldAddEndPuncttrue
\mciteSetBstMidEndSepPunct{\mcitedefaultmidpunct}
{\mcitedefaultendpunct}{\mcitedefaultseppunct}\relax
\EndOfBibitem
\bibitem[Tao \latin{et~al.}(2024)Tao, Qiu, Bhati, Bian, Duston, Rawlinson, Littlejohn, and Subotnik]{coraline24gamma}
Tao,~Z.; Qiu,~T.; Bhati,~M.; Bian,~X.; Duston,~T.; Rawlinson,~J.; Littlejohn,~R.~G.; Subotnik,~J.~E. Practical Phase-Space Electronic Hamiltonians for Ab Initio Dynamics. \textbf{2024}, \relax
\mciteBstWouldAddEndPunctfalse
\mciteSetBstMidEndSepPunct{\mcitedefaultmidpunct}
{}{\mcitedefaultseppunct}\relax
\EndOfBibitem
\bibitem[Bian \latin{et~al.}(2023)Bian, Tao, Wu, Rawlinson, Littlejohn, and Subotnik]{berryforce}
Bian,~X.; Tao,~Z.; Wu,~Y.; Rawlinson,~J.; Littlejohn,~R.; Subotnik,~J. Total Angular Momentum Conservation in Ab Initio Born-Oppenheimer Molecular Dynamics. \emph{Physical Review B} \textbf{2023}, \relax
\mciteBstWouldAddEndPunctfalse
\mciteSetBstMidEndSepPunct{\mcitedefaultmidpunct}
{}{\mcitedefaultseppunct}\relax
\EndOfBibitem
\bibitem[Mai \latin{et~al.}(2015)Mai, Marquetand, and Gonz{\'a}lez]{ISCReview}
Mai,~S.; Marquetand,~P.; Gonz{\'a}lez,~L. A general method to describe intersystem crossing dynamics in trajectory surface hopping. \emph{International Journal of Quantum Chemistry} \textbf{2015}, \emph{115}, 1215--1231\relax
\mciteBstWouldAddEndPuncttrue
\mciteSetBstMidEndSepPunct{\mcitedefaultmidpunct}
{\mcitedefaultendpunct}{\mcitedefaultseppunct}\relax
\EndOfBibitem
\bibitem[Abedi \latin{et~al.}(2010)Abedi, Maitra, and Gross]{abedi:2010:prl_exact_factorization}
Abedi,~A.; Maitra,~N.~T.; Gross,~E. K.~U. Exact Factorization of the Time-Dependent Electron-Nuclear Wave Function. \emph{Physical Review Letters} \textbf{2010}, \emph{105}, 123002\relax
\mciteBstWouldAddEndPuncttrue
\mciteSetBstMidEndSepPunct{\mcitedefaultmidpunct}
{\mcitedefaultendpunct}{\mcitedefaultseppunct}\relax
\EndOfBibitem
\bibitem[Abedi \latin{et~al.}(2012)Abedi, Maitra, and Gross]{abedi:2012:jcp_exact_factorization}
Abedi,~A.; Maitra,~N.~T.; Gross,~E. K.~U. Correlated electron-nuclear dynamics: Exact factorization of the molecular wavefunction. \emph{Journal of Chemical Physics} \textbf{2012}, \emph{137}, 22A530\relax
\mciteBstWouldAddEndPuncttrue
\mciteSetBstMidEndSepPunct{\mcitedefaultmidpunct}
{\mcitedefaultendpunct}{\mcitedefaultseppunct}\relax
\EndOfBibitem
\bibitem[Ibele \latin{et~al.}(2023)Ibele, Sangiogo~Gil, Curchod, and Agostini]{agostini:2023:jpcl:ci}
Ibele,~L.~M.; Sangiogo~Gil,~E.; Curchod,~B.~F.; Agostini,~F. On the Nature of Geometric and Topological Phases in the Presence of Conical Intersections. \emph{Journal of Physical Chemistry Letters} \textbf{2023}, \emph{14}, 11625--11631\relax
\mciteBstWouldAddEndPuncttrue
\mciteSetBstMidEndSepPunct{\mcitedefaultmidpunct}
{\mcitedefaultendpunct}{\mcitedefaultseppunct}\relax
\EndOfBibitem
\bibitem[Arribas and Maitra(2024)Arribas, and Maitra]{maitra:2024:electronic_coherences_exact_factorization}
Arribas,~E.~V.; Maitra,~N.~T. Electronic Coherences in Molecules: The Projected Nuclear Quantum Momentum as a Hidden Agent. 2024\relax
\mciteBstWouldAddEndPuncttrue
\mciteSetBstMidEndSepPunct{\mcitedefaultmidpunct}
{\mcitedefaultendpunct}{\mcitedefaultseppunct}\relax
\EndOfBibitem
\bibitem[Curchod \latin{et~al.}(2016)Curchod, Agostini, and Gross]{curchod:2016:exact_factorization}
Curchod,~B. F.~E.; Agostini,~F.; Gross,~E. K.~U. {An exact factorization perspective on quantum interferences in nonadiabatic dynamics}. \emph{Journal of Chemical Physics} \textbf{2016}, \emph{145}, 034103\relax
\mciteBstWouldAddEndPuncttrue
\mciteSetBstMidEndSepPunct{\mcitedefaultmidpunct}
{\mcitedefaultendpunct}{\mcitedefaultseppunct}\relax
\EndOfBibitem
\bibitem[Li \latin{et~al.}(2022)Li, Requist, and Gross]{li_gross:2022:PRL:angular_momentum_transfer}
Li,~C.; Requist,~R.; Gross,~E. Energy, Momentum, and Angular Momentum Transfer between Electrons and Nuclei. \emph{Physical Review Letters} \textbf{2022}, \emph{128}, 113001\relax
\mciteBstWouldAddEndPuncttrue
\mciteSetBstMidEndSepPunct{\mcitedefaultmidpunct}
{\mcitedefaultendpunct}{\mcitedefaultseppunct}\relax
\EndOfBibitem
\bibitem[Tully(1990)]{Tully1990}
Tully,~J.~C. Molecular dynamics with electronic transitions. \emph{The Journal of Chemical Physics} \textbf{1990}, \emph{93}, 1061--1071\relax
\mciteBstWouldAddEndPuncttrue
\mciteSetBstMidEndSepPunct{\mcitedefaultmidpunct}
{\mcitedefaultendpunct}{\mcitedefaultseppunct}\relax
\EndOfBibitem
\bibitem[Kapral and Ciccotti(1999)Kapral, and Ciccotti]{kapral1999}
Kapral,~R.; Ciccotti,~G. {Mixed quantum-classical dynamics}. \emph{The Journal of Chemical Physics} \textbf{1999}, \emph{110}, 8919--8929\relax
\mciteBstWouldAddEndPuncttrue
\mciteSetBstMidEndSepPunct{\mcitedefaultmidpunct}
{\mcitedefaultendpunct}{\mcitedefaultseppunct}\relax
\EndOfBibitem
\bibitem[Peng \latin{et~al.}(2019)Peng, Xie, Hu, and Lan]{adiabatic}
Peng,~J.; Xie,~Y.; Hu,~D.; Lan,~Z. {Performance of trajectory surface hopping method in the treatment of ultrafast intersystem crossing dynamics}. \emph{The Journal of Chemical Physics} \textbf{2019}, \emph{150}, 164126\relax
\mciteBstWouldAddEndPuncttrue
\mciteSetBstMidEndSepPunct{\mcitedefaultmidpunct}
{\mcitedefaultendpunct}{\mcitedefaultseppunct}\relax
\EndOfBibitem
\bibitem[Shenvi(2009)]{shenviPSSH}
Shenvi,~N. {Phase-space surface hopping: Nonadiabatic dynamics in a superadiabatic basis}. \emph{The Journal of Chemical Physics} \textbf{2009}, \emph{130}, 124117\relax
\mciteBstWouldAddEndPuncttrue
\mciteSetBstMidEndSepPunct{\mcitedefaultmidpunct}
{\mcitedefaultendpunct}{\mcitedefaultseppunct}\relax
\EndOfBibitem
\bibitem[Gherib \latin{et~al.}(2016)Gherib, Ye, Ryabinkin, and Izmaylov]{izmaylov2016jpc_dboc_pssh}
Gherib,~R.; Ye,~L.; Ryabinkin,~I.~G.; Izmaylov,~A.~F. On the inclusion of the diagonal Born-Oppenheimer correction in surface hopping methods. \emph{Journal of Chemical Physics} \textbf{2016}, \emph{144}, 154103\relax
\mciteBstWouldAddEndPuncttrue
\mciteSetBstMidEndSepPunct{\mcitedefaultmidpunct}
{\mcitedefaultendpunct}{\mcitedefaultseppunct}\relax
\EndOfBibitem
\bibitem[Athavale \latin{et~al.}(2023)Athavale, Bian, Tao, Wu, Qiu, Rawlinson, Littlejohn, and Subotnik]{athavale2023}
Athavale,~V.; Bian,~X.; Tao,~Z.; Wu,~Y.; Qiu,~T.; Rawlinson,~J.; Littlejohn,~R.~G.; Subotnik,~J.~E. Surface hopping, electron translation factors, electron rotation factors, momentum conservation, and size consistency. \emph{The Journal of Chemical Physics} \textbf{2023}, \emph{159}\relax
\mciteBstWouldAddEndPuncttrue
\mciteSetBstMidEndSepPunct{\mcitedefaultmidpunct}
{\mcitedefaultendpunct}{\mcitedefaultseppunct}\relax
\EndOfBibitem
\bibitem[{Lengsfield} \latin{et~al.}(1984){Lengsfield}, {Saxe}, and {Yarkony}]{yarkony}
{Lengsfield},~I.,~Byron~H.; {Saxe},~P.; {Yarkony},~D.~R. {On the evaluation of nonadiabatic coupling matrix elements using SA-MCSCF/CI wave functions and analytic gradient methods. I}. \emph{The Journal of Chemical Physics} \textbf{1984}, \emph{81}, 4549--4553\relax
\mciteBstWouldAddEndPuncttrue
\mciteSetBstMidEndSepPunct{\mcitedefaultmidpunct}
{\mcitedefaultendpunct}{\mcitedefaultseppunct}\relax
\EndOfBibitem
\bibitem[Qiu \latin{et~al.}(2024)Qiu, Bhati, Tao, Bian, Rawlinson, Littlejohn, and Subotnik]{tian24gamma}
Qiu,~T.; Bhati,~M.; Tao,~Z.; Bian,~X.; Rawlinson,~J.; Littlejohn,~R.~G.; Subotnik,~J.~E. A Simple One-Electron Expression for Electron Rotational Factors. \textbf{2024}, \relax
\mciteBstWouldAddEndPunctfalse
\mciteSetBstMidEndSepPunct{\mcitedefaultmidpunct}
{}{\mcitedefaultseppunct}\relax
\EndOfBibitem
\bibitem[Littlejohn \latin{et~al.}(2023)Littlejohn, Rawlinson, and Subotnik]{LittlejohnRawlinsonSubotnik23}
Littlejohn,~R.; Rawlinson,~J.; Subotnik,~J. Representation and conservation of angular momentum in the {B}orn-{O}ppenheimer theory of polyatomic molecules. \emph{J. Chem. Phys.} \textbf{2023}, \emph{158}, 104302\relax
\mciteBstWouldAddEndPuncttrue
\mciteSetBstMidEndSepPunct{\mcitedefaultmidpunct}
{\mcitedefaultendpunct}{\mcitedefaultseppunct}\relax
\EndOfBibitem
\bibitem[Littlejohn \latin{et~al.}(2024)Littlejohn, Rawlinson, and Subotnik]{LittlejohnRawlinsonSubotnik24}
Littlejohn,~R.; Rawlinson,~J.; Subotnik,~J. Diagonalizing the {B}orn-{O}ppenheimer {H}amiltonian via {M}oyal perturbation theory, nonadiabatic corrections, and translational degrees of freedom. \emph{J. Chem. Phys.} \textbf{2024}, \emph{160}, 114103\relax
\mciteBstWouldAddEndPuncttrue
\mciteSetBstMidEndSepPunct{\mcitedefaultmidpunct}
{\mcitedefaultendpunct}{\mcitedefaultseppunct}\relax
\EndOfBibitem
\bibitem[Power and Thirunamachandran(1974)Power, and Thirunamachandran]{power}
Power,~E.~A.; Thirunamachandran,~T. {Circular dichroism: A general theory based on quantum electrodynamics}. \emph{The Journal of Chemical Physics} \textbf{1974}, \emph{60}, 3695--3701\relax
\mciteBstWouldAddEndPuncttrue
\mciteSetBstMidEndSepPunct{\mcitedefaultmidpunct}
{\mcitedefaultendpunct}{\mcitedefaultseppunct}\relax
\EndOfBibitem
\bibitem[Shao \latin{et~al.}(2015)Shao, Gan, Epifanovsky, Gilbert, Wormit, Kussmann, Lange, Behn, Deng, Feng, Ghosh, Goldey, Horn, Jacobson, Kaliman, Khaliullin, Kuś, Landau, Liu, Proynov, Rhee, Richard, Rohrdanz, Steele, Sundstrom, III, Zimmerman, Zuev, Albrecht, Alguire, Austin, Beran, Bernard, Berquist, Brandhorst, Bravaya, Brown, Casanova, Chang, Chen, Chien, Closser, Crittenden, Diedenhofen, Jr., Do, Dutoi, Edgar, Fatehi, Fusti-Molnar, Ghysels, Golubeva-Zadorozhnaya, Gomes, Hanson-Heine, Harbach, Hauser, Hohenstein, Holden, Jagau, Ji, Kaduk, Khistyaev, Kim, Kim, King, Klunzinger, Kosenkov, Kowalczyk, Krauter, Lao, Laurent, Lawler, Levchenko, Lin, Liu, Livshits, Lochan, Luenser, Manohar, Manzer, Mao, Mardirossian, Marenich, Maurer, Mayhall, Neuscamman, Oana, Olivares-Amaya, O’Neill, Parkhill, Perrine, Peverati, Prociuk, Rehn, Rosta, Russ, Sharada, Sharma, Small, Sodt, Stein, Stück, Su, Thom, Tsuchimochi, Vanovschi, Vogt, Vydrov, Wang, Watson, Wenzel, White, Williams, Yang, Yeganeh, Yost, You, Zhang,
  Zhang, Zhao, Brooks, Chan, Chipman, Cramer, III, Gordon, Hehre, Klamt, III, Schmidt, Sherrill, Truhlar, Warshel, Xu, Aspuru-Guzik, Baer, Bell, Besley, Chai, Dreuw, Dunietz, Furlani, Gwaltney, Hsu, Jung, Kong, Lambrecht, Liang, Ochsenfeld, Rassolov, Slipchenko, Subotnik, Voorhis, Herbert, Krylov, Gill, and Head-Gordon]{qchem4}
Shao,~Y.; Gan,~Z.; Epifanovsky,~E.; Gilbert,~A.~T.; Wormit,~M.; Kussmann,~J.; Lange,~A.~W.; Behn,~A.; Deng,~J.; Feng,~X.; Ghosh,~D.; Goldey,~M.; Horn,~P.~R.; Jacobson,~L.~D.; Kaliman,~I.; Khaliullin,~R.~Z.; Kuś,~T.; Landau,~A.; Liu,~J.; Proynov,~E.~I.; Rhee,~Y.~M.; Richard,~R.~M.; Rohrdanz,~M.~A.; Steele,~R.~P.; Sundstrom,~E.~J.; III,~H. L.~W.; Zimmerman,~P.~M.; Zuev,~D.; Albrecht,~B.; Alguire,~E.; Austin,~B.; Beran,~G. J.~O.; Bernard,~Y.~A.; Berquist,~E.; Brandhorst,~K.; Bravaya,~K.~B.; Brown,~S.~T.; Casanova,~D.; Chang,~C.-M.; Chen,~Y.; Chien,~S.~H.; Closser,~K.~D.; Crittenden,~D.~L.; Diedenhofen,~M.; Jr.,~R. A.~D.; Do,~H.; Dutoi,~A.~D.; Edgar,~R.~G.; Fatehi,~S.; Fusti-Molnar,~L.; Ghysels,~A.; Golubeva-Zadorozhnaya,~A.; Gomes,~J.; Hanson-Heine,~M.~W.; Harbach,~P.~H.; Hauser,~A.~W.; Hohenstein,~E.~G.; Holden,~Z.~C.; Jagau,~T.-C.; Ji,~H.; Kaduk,~B.; Khistyaev,~K.; Kim,~J.; Kim,~J.; King,~R.~A.; Klunzinger,~P.; Kosenkov,~D.; Kowalczyk,~T.; Krauter,~C.~M.; Lao,~K.~U.; Laurent,~A.~D.; Lawler,~K.~V.;
  Levchenko,~S.~V.; Lin,~C.~Y.; Liu,~F.; Livshits,~E.; Lochan,~R.~C.; Luenser,~A.; Manohar,~P.; Manzer,~S.~F.; Mao,~S.-P.; Mardirossian,~N.; Marenich,~A.~V.; Maurer,~S.~A.; Mayhall,~N.~J.; Neuscamman,~E.; Oana,~C.~M.; Olivares-Amaya,~R.; O’Neill,~D.~P.; Parkhill,~J.~A.; Perrine,~T.~M.; Peverati,~R.; Prociuk,~A.; Rehn,~D.~R.; Rosta,~E.; Russ,~N.~J.; Sharada,~S.~M.; Sharma,~S.; Small,~D.~W.; Sodt,~A.; Stein,~T.; Stück,~D.; Su,~Y.-C.; Thom,~A.~J.; Tsuchimochi,~T.; Vanovschi,~V.; Vogt,~L.; Vydrov,~O.; Wang,~T.; Watson,~M.~A.; Wenzel,~J.; White,~A.; Williams,~C.~F.; Yang,~J.; Yeganeh,~S.; Yost,~S.~R.; You,~Z.-Q.; Zhang,~I.~Y.; Zhang,~X.; Zhao,~Y.; Brooks,~B.~R.; Chan,~G.~K.; Chipman,~D.~M.; Cramer,~C.~J.; III,~W. A.~G.; Gordon,~M.~S.; Hehre,~W.~J.; Klamt,~A.; III,~H. F.~S.; Schmidt,~M.~W.; Sherrill,~C.~D.; Truhlar,~D.~G.; Warshel,~A.; Xu,~X.; Aspuru-Guzik,~A.; Baer,~R.; Bell,~A.~T.; Besley,~N.~A.; Chai,~J.-D.; Dreuw,~A.; Dunietz,~B.~D.; Furlani,~T.~R.; Gwaltney,~S.~R.; Hsu,~C.-P.; Jung,~Y.; Kong,~J.;
  Lambrecht,~D.~S.; Liang,~W.; Ochsenfeld,~C.; Rassolov,~V.~A.; Slipchenko,~L.~V.; Subotnik,~J.~E.; Voorhis,~T.~V.; Herbert,~J.~M.; Krylov,~A.~I.; Gill,~P.~M.; Head-Gordon,~M. Advances in molecular quantum chemistry contained in the Q-Chem 4 program package. \emph{Molecular Physics} \textbf{2015}, \emph{113}, 184--215\relax
\mciteBstWouldAddEndPuncttrue
\mciteSetBstMidEndSepPunct{\mcitedefaultmidpunct}
{\mcitedefaultendpunct}{\mcitedefaultseppunct}\relax
\EndOfBibitem
\bibitem[Stephens \latin{et~al.}(1994)Stephens, Chabalowski, Devlin, and Jalkanen]{expvcdcycp}
Stephens,~P.; Chabalowski,~C.; Devlin,~F.; Jalkanen,~K. Ab initio calculation of vibrational circular dichroism spectra using large basis set MP2 force fields. \emph{Chemical physics letters} \textbf{1994}, \emph{225}, 247--257\relax
\mciteBstWouldAddEndPuncttrue
\mciteSetBstMidEndSepPunct{\mcitedefaultmidpunct}
{\mcitedefaultendpunct}{\mcitedefaultseppunct}\relax
\EndOfBibitem
\bibitem[Freedman \latin{et~al.}(1991)Freedman, Spencer, Ragunathan, Nafie, Moore, and Schwab]{expvcdoxi}
Freedman,~T.~B.; Spencer,~K.~M.; Ragunathan,~N.; Nafie,~L.~A.; Moore,~J.~A.; Schwab,~J.~M. Vibrational circular dichroism of (S, S)-[2, 3-2H2] oxirane in the gas phase and in solution. \emph{Canadian journal of chemistry} \textbf{1991}, \emph{69}, 1619--1629\relax
\mciteBstWouldAddEndPuncttrue
\mciteSetBstMidEndSepPunct{\mcitedefaultmidpunct}
{\mcitedefaultendpunct}{\mcitedefaultseppunct}\relax
\EndOfBibitem
\bibitem[Kawiecki \latin{et~al.}(1991)Kawiecki, Devlin, Stephens, and Amos]{expvcdmeoxi}
Kawiecki,~R.; Devlin,~F.; Stephens,~P.; Amos,~R. Vibrational circular dichroism of propylene oxide. \emph{The Journal of Physical Chemistry} \textbf{1991}, \emph{95}, 9817--9831\relax
\mciteBstWouldAddEndPuncttrue
\mciteSetBstMidEndSepPunct{\mcitedefaultmidpunct}
{\mcitedefaultendpunct}{\mcitedefaultseppunct}\relax
\EndOfBibitem
\bibitem[Kubelka \latin{et~al.}(2005)Kubelka, Huang, and Keiderling]{VCDsolvent}
Kubelka,~J.; Huang,~R.; Keiderling,~T.~A. Solvent effects on IR and VCD spectra of helical peptides: DFT-based static spectral simulations with explicit water. \emph{The Journal of Physical Chemistry B} \textbf{2005}, \emph{109}, 8231--8243\relax
\mciteBstWouldAddEndPuncttrue
\mciteSetBstMidEndSepPunct{\mcitedefaultmidpunct}
{\mcitedefaultendpunct}{\mcitedefaultseppunct}\relax
\EndOfBibitem
\bibitem[Poopari \latin{et~al.}(2012)Poopari, Zhu, Dezhahang, and Xu]{VCDHbond}
Poopari,~M.~R.; Zhu,~P.; Dezhahang,~Z.; Xu,~Y. Vibrational absorption and vibrational circular dichroism spectra of leucine in water under different pH conditions: hydrogen-bonding interactions with water. \emph{The Journal of Chemical Physics} \textbf{2012}, \emph{137}\relax
\mciteBstWouldAddEndPuncttrue
\mciteSetBstMidEndSepPunct{\mcitedefaultmidpunct}
{\mcitedefaultendpunct}{\mcitedefaultseppunct}\relax
\EndOfBibitem
\bibitem[Fus{\`e} \latin{et~al.}(2022)Fus{\`e}, Longhi, Mazzeo, Stranges, Leonelli, Aquila, Bodo, Brunetti, Bicchi, Cagliero, \latin{et~al.} others]{fuse2022anharmonic}
Fus{\`e},~M.; Longhi,~G.; Mazzeo,~G.; Stranges,~S.; Leonelli,~F.; Aquila,~G.; Bodo,~E.; Brunetti,~B.; Bicchi,~C.; Cagliero,~C.; others Anharmonic Aspects in Vibrational Circular Dichroism Spectra from 900 to 9000 cm--1 for Methyloxirane and Methylthiirane. \emph{The Journal of Physical Chemistry A} \textbf{2022}, \emph{126}, 6719--6733\relax
\mciteBstWouldAddEndPuncttrue
\mciteSetBstMidEndSepPunct{\mcitedefaultmidpunct}
{\mcitedefaultendpunct}{\mcitedefaultseppunct}\relax
\EndOfBibitem
\bibitem[Barone \latin{et~al.}(2014)Barone, Biczysko, Bloino, and Puzzarini]{assignmodemeoxi}
Barone,~V.; Biczysko,~M.; Bloino,~J.; Puzzarini,~C. Accurate molecular structures and infrared spectra of trans-2, 3-dideuterooxirane, methyloxirane, and trans-2, 3-dimethyloxirane. \emph{The Journal of chemical physics} \textbf{2014}, \emph{141}\relax
\mciteBstWouldAddEndPuncttrue
\mciteSetBstMidEndSepPunct{\mcitedefaultmidpunct}
{\mcitedefaultendpunct}{\mcitedefaultseppunct}\relax
\EndOfBibitem
\bibitem[Kutzelnigg(2007)]{kutzelnigg:2007:mp}
Kutzelnigg,~W. Which masses are vibrating or rotating in a molecule? \emph{Molecular Physics} \textbf{2007}, \emph{105}, 2627--2647\relax
\mciteBstWouldAddEndPuncttrue
\mciteSetBstMidEndSepPunct{\mcitedefaultmidpunct}
{\mcitedefaultendpunct}{\mcitedefaultseppunct}\relax
\EndOfBibitem
\bibitem[Scherrer \latin{et~al.}(2017)Scherrer, Agostini, Sebastiani, Gross, and Vuilleumier]{gross:2017:prx:born_oppenheimer_mass}
Scherrer,~A.; Agostini,~F.; Sebastiani,~D.; Gross,~E. K.~U.; Vuilleumier,~R. On the Mass of Atoms in Molecules: Beyond the Born-Oppenheimer Approximation. \emph{Physical Review X} \textbf{2017}, \emph{7}, 031035\relax
\mciteBstWouldAddEndPuncttrue
\mciteSetBstMidEndSepPunct{\mcitedefaultmidpunct}
{\mcitedefaultendpunct}{\mcitedefaultseppunct}\relax
\EndOfBibitem
\bibitem[Maxwell(1950)]{isotopeSC1}
Maxwell,~E. Isotope Effect in the Superconductivity of Mercury. \emph{Phys. Rev.} \textbf{1950}, \emph{78}, 477--477\relax
\mciteBstWouldAddEndPuncttrue
\mciteSetBstMidEndSepPunct{\mcitedefaultmidpunct}
{\mcitedefaultendpunct}{\mcitedefaultseppunct}\relax
\EndOfBibitem
\bibitem[Reynolds \latin{et~al.}(1950)Reynolds, Serin, Wright, and Nesbitt]{isotopeSC2}
Reynolds,~C.~A.; Serin,~B.; Wright,~W.~H.; Nesbitt,~L.~B. Superconductivity of Isotopes of Mercury. \emph{Phys. Rev.} \textbf{1950}, \emph{78}, 487--487\relax
\mciteBstWouldAddEndPuncttrue
\mciteSetBstMidEndSepPunct{\mcitedefaultmidpunct}
{\mcitedefaultendpunct}{\mcitedefaultseppunct}\relax
\EndOfBibitem
\bibitem[Gao \latin{et~al.}(2023)Gao, Pan, Zhou, and Zhang]{Gao2023}
Gao,~Y.; Pan,~Y.; Zhou,~J.; Zhang,~L. Chiral phonon mediated high-temperature superconductivity. \emph{Physical Review B} \textbf{2023}, \emph{108}, 064510\relax
\mciteBstWouldAddEndPuncttrue
\mciteSetBstMidEndSepPunct{\mcitedefaultmidpunct}
{\mcitedefaultendpunct}{\mcitedefaultseppunct}\relax
\EndOfBibitem
\bibitem[Stephens(1987)]{stephens1987gauge}
Stephens,~P. Gauge dependence of vibrational magnetic dipole transition moments and rotational strengths. \emph{Journal of Physical Chemistry} \textbf{1987}, \emph{91}, 1712--1715\relax
\mciteBstWouldAddEndPuncttrue
\mciteSetBstMidEndSepPunct{\mcitedefaultmidpunct}
{\mcitedefaultendpunct}{\mcitedefaultseppunct}\relax
\EndOfBibitem
\bibitem[Pople \latin{et~al.}(2009)Pople, Krishnan, Schlegel, and Binkley]{MODeriv}
Pople,~J.~A.; Krishnan,~R.; Schlegel,~H.~B.; Binkley,~J.~S. Derivative studies in hartree-fock and m{\o}ller-plesset theories. \emph{International Journal of Quantum Chemistry} \textbf{2009}, \emph{16}, 225--241\relax
\mciteBstWouldAddEndPuncttrue
\mciteSetBstMidEndSepPunct{\mcitedefaultmidpunct}
{\mcitedefaultendpunct}{\mcitedefaultseppunct}\relax
\EndOfBibitem
\end{mcitethebibliography}

\end{document}